\documentclass[aps,prr,superscriptaddress,twocolumn,notitlepage]{revtex4-1}

\usepackage[latin1]{inputenc}
\usepackage{amsmath}
\usepackage{amsfonts}
\usepackage{amssymb}
\usepackage{graphicx}
\usepackage{epstopdf}
\graphicspath{ {images/} }
\usepackage[titletoc]{appendix}
\usepackage{mathrsfs}
\usepackage{subfigure}

\usepackage{graphicx}
\usepackage{soul}
\usepackage{verbatim}

\usepackage[colorlinks=true,citecolor=blue,linkcolor=magenta]{hyperref}

\usepackage{dsfont}
\usepackage{changepage}

\usepackage{amsthm, amssymb}
\usepackage{array}
\usepackage[usenames,dvipsnames]{color}
\usepackage[table]{xcolor}

\def \ua{{\uparrow}}
\def \da{{\downarrow}}
\def \be{\begin{equation}}
\def \ee{\end{equation}}
\def \ba{\begin{array}}
	\def \ea{\end{array}}
\def \bea{\begin{eqnarray}}
\def \eea{\end{eqnarray}}
\def \nn{\nonumber}
\def \ve {\varepsilon}

\def \bk{{\bf k}}

\def \W{{\Omega}}
\def \e{{\epsilon}}

\def \a{{\alpha}}
\def \t{{\theta}}
\def \b{{\beta}}
\def \g{{\gamma}}
\def \D{{\Delta}}
\def \d{{\delta}}
\def \w{{\omega}}
\def \s{{\sigma}}

\def \e{{\epsilon}}
\def \ve{{\varepsilon}}

\def \G{{\Gamma}}
\def \z{{\zeta}}

\def \ba{\begin{align*}}
\def \ea{\end{align*}}

\newcounter{indice}

\def \mrm{\mathrm}

\def \bs{\boldsymbol}
\def \mc{\mathcal}

\def \md{\mathds}

\definecolor{p2}{rgb}{0.8, 0.25, 0.87}
\definecolor{p53}{rgb}{0.932, 0.456, 0.8096}
\definecolor{p1}{rgb}{0.945, 0.7, 0.874}
\definecolor{p13}{rgb}{0.983, 0.884, 0.9524}

\newcommand{\sign}{\rm{sign}}
\newcommand{\Tr}{\rm{Tr}}
\begin{document}
	\title{The effect of interorbital scattering on superconductivity in doped Dirac semimetals}
	\author{David Dentelski}

\affiliation{Department of Physics, Bar-Ilan University, 52900, Ramat Gan Israel}
\affiliation{Center for Quantum Entanglement Science and Technology, Bar-Ilan University, 52900, Ramat Gan Israel}
	\author{Vladyslav Kozii}
\affiliation{Department of Physics, University of California, Berkeley, CA 94720, USA}
\affiliation{Materials Sciences Division, Lawrence Berkeley National Laboratory, Berkeley, CA 94720, USA}
	\author{Jonathan Ruhman$^{1,\,2}$}

\begin{abstract}
Unconventional superconductivity has been discovered in a variety of doped quantum materials, including topological insulators and semimetals. A unifying property of these systems is strong orbital hybridization, which leads to pairing of states with non-trivial Bloch wave functions. 
In contrast to naive expectation, however, many of these superconductors are relatively resilient to disorder.
Here we study the interplay of superconductivity and disorder in doped three-dimensional Dirac systems, which serve as a paradigmatic dispersion in quantum materials, using Abrikosov-Gor'kov theory. 
In this way, the role of disorder is captured by a single parameter $\Gamma$, the pair scattering rate.
In contrast to previous studies, we argue that interorbital scattering can not be neglected due to the strong orbital hybridization in Dirac systems. 
We find that  the robustness of different pairing states highly depends on the relative strength of the different interorbital scattering channels.
In particular, we find that the ``nematic'' superconducting state, which is argued to be the ground state in many Bi$_2$Se$_3$ related compounds, is not protected from disorder in any way. The pair scattering rate in this case is at best smaller by a factor of 3 compared to systems without spin-orbit coupling.
We also find that the odd-parity pairing state with total angular momentum zero (the B-phase of superfluid $^3$He) is protected against certain types of disorder, which include a family of time-reversal odd (magnetic) impurities. 
Namely, this odd-pairty state is a singlet of partners under $\mc{CT}$ symmetry (rather than $\mc T$ symmetry in the standard Anderson's theory), where $\mc C$ and $\mc T$ are chiral and time-reversal symmetries, respectively. As a result, it is protected against any disorder potential that respects $\mc{CT}$ symmetry. 
Our procedure is very general and can be readily applied to different band structures and disorder configurations.  
\end{abstract}
	
\maketitle

\section{Introduction}
Anderson's theory explains why conventional s-wave superconductors are weakly affected by non-magnetic disorder~\cite{anderson1959theory,abrikosov1960contribution,Gorkov2008}. It is based on two essential conditions. The first one is that the Cooper pairs in these superconductors form singlet states of time-reversed partners. The second one is that the phase of the pair wave function is featureless over the entire Fermi surface. These two ensure that the pairing interaction, written in the basis that diagonalizes the disorder potential, remains the same as in the clean limit.  Consequently, one can always pair time-reversed partners with the same interaction and the same transition temperature~\cite{ma1985localized}.  
In contrast, the Cooper pairs in unconventional superconductors violate one of these conditions and, as a result, are not protected~\cite{Larkin1965,millis1988inelastic,radtke1993predictions,hirschfeld1993effect,emery1995importance,dalichaouch1995impurity,mackenzie1998extremely,mackenzie2003superconductivity,fujita2005effect,florens2005impact,alloul2009defects,tarantini2010suppression,li2012superconductivity,kirshenbaum2012universal,mizukami2014disorder,bovzovic2016dependence,lee2017disorder,khestanova2018unusual}.

From the theoretical perspective, the conditions to prefer pairing in non-$s$-wave channels are quite stringent, even without the destructive effect of disorder. 
Nonetheless, a large body of recent experimental measurements, performed in doped topological materials, is consistent with an unconventional superconducting state~\cite{hor2010superconductivity,wray2010observation,butch2011superconductivity,sasaki2012odd,wang2016observation,matano2016spin,yonezawa2017thermodynamic,WillaWillaetal2018,PandeVisseretal2016,NikitindeVisseretal2016,AsabaLuLietal2017,Taoetal2018,shen2017nematic,kim2018beyond}, which was predicted theoretically~\cite{fu2010odd,sato2010,Funematic2014,kozii2015odd,brydon2016pairing,savary2017superconductivity}. 
Surprisingly, these superconductors are extremely robust to disorder~\cite{kriener2012anomalous,novak2013unusual,smylie2017robust,andersen2020generalized,timmons2020electron}.

\begin{figure} [h!]
	\centering
	\includegraphics[width=0.5\textwidth]{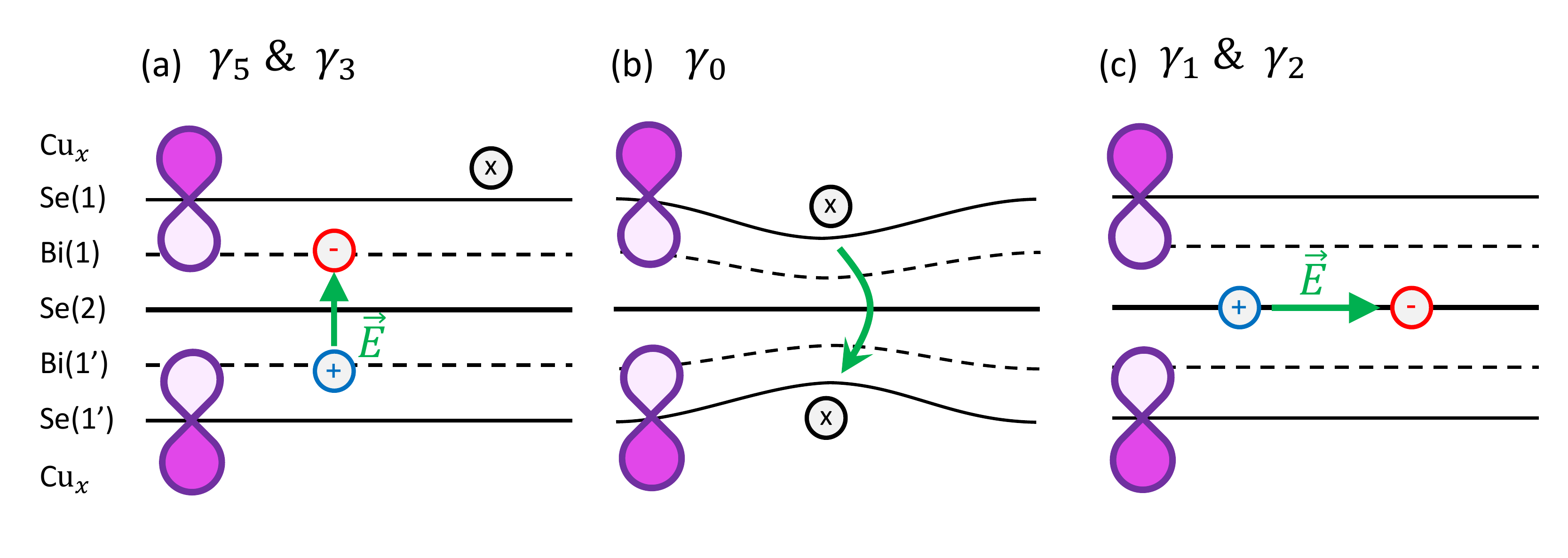} 
	\caption{Different types of non-trivial time-reversal-symmetric intra- and interorbital disorder potentials in Bi$_2$Se$_3$ (as a prototypical Dirac system). The solid and dashed lines represent the Se and Bi layers in the quintuple unit cell, respectively. The purple $p_z$ orbitals represent the itinerant states on the top and bottom Se layers, which disperse as Dirac fermions close to the $\Gamma$-point. (a) Disorder that breaks the symmetry between top and bottom layers induces $\gamma_5$ and $\gamma_3$ potentials [our convention for the $\gamma$ matrices is given below Eq.\eqref{eq:S0}]. This can be caused by a polar impurity or a charged impurity closer to one layer than the other (e.g., due to the intercalation of Cu). (b) Disorder that modifies the tunneling between top and bottom (either by causing a barrier or by stretching/squeezing the $z$-axis) induces disorder in the mass term of the form $\gamma_0$. (c) An in-plane polar impurity induces $\gamma_1$ and $\gamma_2$ potentials. We note that in all cases we also anticipate an intraorbital density potential (proportional to the identity matrix). }
	\label{fig:structure}
\end{figure}

Mechanisms based on the huge spin-orbit coupling characterizing the topological materials have been suggested to explain this robustness.
The authors of Ref.~\cite{michaeli2012spin} studied the effect of disorder on an odd-parity paring state with zero total angular momentum (equivalent to the B-phase in superfluid $^3$He). They found that an additional chiral symmetry can protect this state from certain types disorder when it is present.
It was later suggested that the pair wavefunction in doped Bi$_2$Se$_3$ is a multi-component nematic state {
 which breaks the rotational symmetry of the crystal~\cite{Funematic2014,VenderbosKoziiFu2016}.  References~\cite{nagai2015robust,andersen2020generalized} studied the effect of disorder on the nodal nematic state, also arguing for some robustness, however, their results were based on less generic grounds. Studies of the effects of scalar disorder on unconventional paring in topological materials were recently discussed for materials other than Bi$_2$Se$_3$~\cite{cavanagh2020robustness,timmons2020electron}, and also in the context of the surface of topological materials~\cite{ito2011stability, ito2012impurity, tkachov2013suppression, ozfidan2016gapless}}.
 

The studies mentioned above, however, focus only on the effects of intraorbital scattering, which has equal weight on all orbitals (i.e., density disorder). Because topological materials are  multiorbital systems with huge orbital hybridization, interorbital scattering is not expected to be particularly weaker than density disorder. In Fig.~\ref{fig:structure} we schematically depict three types of time-reversal-symmetric disorder potentials, which are expected to be present in Bi$_2$Se$_3$ and lead to interorbital scattering. Thus, it is important to understand the effects of interorbital scattering in the superconducting topological materials.

The influence of interorbital scattering on pairing was first emphasized by Golubov and Mazin~\cite{golubov1997effect} and was later studied in the context of systems with multiple Fermi surfaces (see, for example, Refs.~\cite{zhang2009orbital,efremov2011disorder,wang2013using,trevisan2018impact,mockli2018robust}).  We emphasize that in topological materials the multiorbital nature is embedded in the Bloch wave-functions rather than the presence of multiple Fermi surfaces, making them somewhat different. 

In this paper, we study the effect of short-ranged intra- and interorbital scattering on superconductivity in  three-dimensional materials with {Dirac dispersion}. The Dirac dispersion is a paradigmatic example of dispersion relations in topological materials, which also naturally have large spin-orbit coupling~\cite{ando2015topological}.

{ We provide an extensive picture of how the transition temperature $T_c$ in different pairing channels is affected by all possible types of short-ranged scattering potentials. 
Our results are expressed in terms of the pair-breaking rate $\Gamma$, which also enters the Abrikosov-Gor'kov theory of superconductors with magnetic impurities~\cite{abrikosov1960contribution}.  
 We first discuss the case of a massless Dirac dispersion where the pair breaking rates are proportional to the single particle scattering rates through universal rational numbers (see Table~\ref{Tbl:rhoc}). We then compute how these scattering rates vary with mass, which allows us to interpolate to the well known results for the systems with no spin-orbit coupling~\cite{abrikosov1960contribution,Larkin1965}.}

We find that the only state robust to time-reversal symmetric (TRS) disorder is the $s$-wave, in agreement with the Anderson's theorem~\cite{anderson1959theory}. Another fully gapped isotropic state is the odd-parity state with zero total angular momentum (analogous to the B-phase in superfluid $^3$He), which does not exhibit such robustness, in contrast to previous expectations~\cite{michaeli2012spin,timmons2020electron}. Despite being protected against certain types of disorder, it turns out to be very sensitive to some other types of defects, such as mass and polar impurities, which we expect to be generally present in topological materials.

{On the other hand, we find that this fully isotropic odd-parity state is protected from any disorder that respects $\mc{CT}$ (assuming this symmetry is also present in the clean system), where $\mc C$ and $\mc T$ are chiral and time-reversal symmetries, respectively. The reason for such robustness is that this state  corresponds to a singlet pairing state of $\mc{CT}$ partners, in perfect analogy to pairing of $\mc T$-partners in Anderson's original argument~\cite{anderson1959theory}.  Interestingly, this result implies that the fully gapped odd-parity state is protected against certain disorder potentials that are odd under time-reversal $\mc T$.

We also study the multicomponent states, which are the $O(3)$-symmetry group
related to the nematic pairing states in doped $\text{Bi}_2\text{Se}_3$. 
The realtion between these states is obtained when the symmetry group is reduced from the fully isotropic $O(3)$ to trigonal $D_{3d}$ group of $\text{Bi}_2\text{Se}_3$}. We find that the pair breaking rate for these states can be smaller than the corresponding rate in systems without spin-orbit coupling. However, they are still significantly influenced by disorder in contrast to previous studies~\cite{nagai2015robust,andersen2020generalized}. We also consider the effect of magnetic impurities, where the $O(3)$-symmetry pairing state can also be slightly more protected than in systems without spin-orbit coupling~\cite{Larkin1965} (depending on the microscopic nature of the magnetic impurities). It should be added that when time-reversal is broken the chiral state might be preferred over the nematic one~\cite{yuan2017superconductivty,chirolli2017time,chirolli2018chiral2,chirolli2020surface,ZyuzinGaraudBabaev2017}.


The rest of this paper is organized as follows. In Sec.~\ref{sec:Model} we present the basic ingredients of our model, namely, an action for a Dirac fermion subjected to a generic disorder potential and an attractive pairing interaction. We project the disorder onto the Bloch basis of the conduction electrons near Fermi surface, which we will use throughout this paper. In Sec.~\ref{sec:Tc} we  calculate the pair breaking rate $\G$ in topological materials, which is the main parameter that affects $T_{c}$ and the whole thermodynamics. Our results are presented in Sec.~\ref{sec:Results}, where we discuss the effect of different types of disorder on the various pairing channels.  Finally, Sec.~\ref{sec:Discussion} provides a summary of our main results and a discussion of the application of our method for other systems. Multiple technical details of our calculation are delegated to the Appendices.

\section{The model} \label{sec:Model}
We start by describing the normal-state action of the model. We consider massless Dirac fermions
\begin{align}\label{eq:S0}
\mc S_0 = \sum_{\w,\bs k} \psi_{\w,\bs{k}}^\dag \left[ -i\w + i v \bs k \cdot \g_0 \bs \g+\d \cdot \g_0 -\e_F \right]\psi_{\w,\bs{k}},
\end{align}	
where $\psi^{\dagger}_{\w,\bs{k}} = \begin{pmatrix} \psi^\dagger_{\w,\bs{k}, +} & \psi^\dagger_{\w,\bs{k}, -} \end{pmatrix}$, $\psi_{\omega,\bs k,\pm}$ correspond to two orbitals, each consisting of a Kramers pair. The orbitals are related to each other through inversion.  $v$ is an isotropic velocity and $\e_F$ is the Fermi energy. 
In addition, the $\g$ matrices are taken to be Hermitian, $\bs\g = \tau_2 \bs s$, $\g_0 = \tau_1 s_0$ and $\gamma_{5}=\g_0 \g_1 \g_2 \g_3 = -\tau_3 s_0$, where $\bs \tau$ and $\bs s$ are Pauli matrices in the orbital and spin basis, correspondingly. { $\d$ is the mass of the Dirac point. In the following, we set $\d=0$ for simplicity unless specified explicitly otherwise. Note that we have neglected higher order corrections in momentum. }

The action in Eq.~\eqref{eq:S0} can be conveniently diagonalized in the manifestly covariant Bloch basis (MCBB), in which the electron spinor transforms as an ordinary $SU(2)$ spin-$1/2$~\cite{fu2015parity,kozii2015odd,VenderbosKoziiFu2016,kozii2019superconductivty}:
$
|\hat{\bs k},1,\z \rangle = {1\over 2}
\begin{pmatrix}
\z - \hat k_z, &
-\hat k_+,&
\z+\hat k_z,&
\hat k_+
\end{pmatrix}^T
$
and 
$
|\hat{\bs k},2,\z\rangle = {1\over 2}
\begin{pmatrix}
-\hat k_-, &
\z + \hat k_z,&
\hat k_-,&
\z - \hat k_z
\end{pmatrix}^T
$, 
where $\z =\pm 1$ corresponds to conduction/valence band, respectively, $\hat k_j = k_j / k$, and $\hat k_{\pm} = \hat k_x \pm i \hat k_y$. Without loss of generality we assume electron doping ($\z=1$), and therefore omit index $\z$ henceforth. The field operators are then approximated by their weight on the band operators $\psi_{\bs k} \approx |\hat {\bs k},1\rangle c_{\bs k, 1}+|\hat {\bs k},2\rangle c_{\bs k, 2}$.

Finally, the action in Eq.~\eqref{eq:S0} possesses inversion, chiral, and time-reversal symmetries. The representation of these symmetry operations in orbital basis is given by $\mc I = \g_0$, $\mc C = \g_5$, and $\mc T = \mc K \g_1 \g_3$, respectively, where $\mc K$ is complex conjugation.

Next, we consider the disorder potential. The crucial element in our theory is the inclusion of interorbital scattering. Within the Dirac notations, such  (momentum-independent) scatterings can be represented using the Dirac matrices introduced above and their products. For elastic short-ranged scattering (compared to $k_F$) we have  
\begin{align}\label{eq:Disorder}
\mc S_d = \sum_{m=0}^{15} \sum_{l=1}^{N_{m}}\sum_{\w,\bs k,\bs p} V_{l,m}\, e^{i(\bs k - \bs p)\cdot \bs r_l} \psi_{\w,\bs p}^\dag M_{m} \psi_{\w,\bs k},
\end{align}
where $N_m$ is the number of impurities in channel $m$ and $\bs r_l$ is the position of these short-ranged impurities. { There are 16 different Hermitian matrices $M_m$ representing different types of disorder.}
The representation of these matrices in the orbital-spin $\tau \otimes s$ basis and their discrete symmetry properties are given in Table~\ref{tbl:Q}. 
We further clarify that $m=0$ corresponds to simple density disorder $\md 1$, which was considered in Refs.~\cite{michaeli2012spin,nagai2015robust,andersen2020generalized,cavanagh2020robustness}. 
$m=1$ corresponds to mass disorder $\g_0$, $m=2$ is odd-parity scalar disorder $\g_5$, $m=3,4,5$ correspond to odd-parity dipolar disorder and the magnetic disorder $m=6,7,8$ correspond to a fully symmetric local moment with spin oriented along the axis $S_\a = -i\ve_{\a\b\g}\g_\b \g_\g$.
Some examples of non-trivial scattering matrices of this type, which naturally appear in disordered Bi$_2$Se$_3$, are shown schematically in Fig.~\ref{fig:structure}.
When projecting the disorder potential onto the MCBB, we obtain a set of scattering matrices $Q_m(\hat {\bs p}, \hat {\bs k})$ with non-trivial momentum dependence:  
\begin{align}\label{eq:Disorder_band_basis} 
\mc S_d = \sum_{m=0}^{15} \sum_{l=1}^{N_{m}}\sum_{\w,\bs k,\bs p} V_{l,m}\, e^{i(\bs k - \bs p)\cdot \bs r_l} c_{\w,\bs p}^\dag Q_{m}(\hat{\bs p},\hat{\bs k}) c_{\w,\bs k},
\end{align}
where $c_{\omega, \bs p}^\dag = (c_{\omega, \bs p,1}^\dag,c_{\omega, \bs p,2}^\dag)$, and we defined the matrices $Q_m^{\a\b}(\hat{\bs p},\hat{\bs k}) \equiv \langle \hat{\bs p}\a|M_m|\hat{\bs k}\b\rangle$, which are listed in Table~\ref{tbl:Q}.

\begin{table*}
	\begin{tabular}{ |c|c|c|c|c|c| }
		\hline
		$m$& $\mc I$ & $\mc T$ &$\mc C$ &Orbital Matrix - $M_m$ & Band Matrix - $\langle\hat{\bs p} \z |M_m|\hat{\bs k} \z\rangle$ \\
		\hline
	   0    & + & + & +&$\md 1 = \tau_0s_0$ & $Q_0(\hat{\bs p},\hat{\bs k}) = {1\over 2}\left(1+\hat{\bs p}\cdot\hat{\bs k}+i[\hat {\bs p}\times\hat{\bs k}]\cdot \bs \s \right)$ \\
		\hline
	   1    & + & + & - & $\g_0=\tau_1 s_0$ & $Q_1(\hat{\bs p},\hat{\bs k}) = {1\over 2}\left(1-\hat{\bs p}\cdot\hat{\bs k}-i[\hat {\bs p}\times\hat{\bs k}]\cdot \bs \s \right)$ \\ 
		\hline
		2    & - & + & + & $\g_5 = -\tau_3 s_0$ & $Q_2(\hat{\bs p},\hat{\bs k}) = {\z\over 2}\left( \hat{\bs{p}}+\hat{\bs{k}}\right)\cdot \bs \s$ \\
		\hline
		3    & - & + & - &$\g_1 = \tau_2 s_1$ & $Q_3(\hat{\bs p},\hat{\bs k}) = {\z\over 2}\left\{ i\left( \hat{\bs{p}}-\hat{\bs{k}}\right)_x + [(\hat{\bs p}+\hat{\bs k})\times \bs \s]_x \right\}$ \\
		\hline
		4    & - & + & - &$\g_2 = \tau_2 s_2$ & $Q_4(\hat{\bs p},\hat{\bs k}) = {\z\over 2}\left\{ i\left( \hat{\bs{p}}-\hat{\bs{k}}\right)_y + [(\hat{\bs p}+\hat{\bs k})\times \bs \s]_y \right\}$ \\
		\hline
		5    & - & + & - &$\g_3 = \tau_2 s_3$ & $Q_5(\hat{\bs p},\hat{\bs k}) = {\z\over 2}\left\{ i\left( \hat{\bs{p}}-\hat{\bs{k}}\right)_z + [(\hat{\bs p}+\hat{\bs k})\times \bs \s]_z \right\}$ \\
		\hline
	    	6    & + & - & + &$i\g_3 \g_2{ = \tau_0 s_1}$ & $Q_6(\hat{\bs p},\hat{\bs k}) = {1\over 2}\left\{ -i [\hat{\bs p}\times \hat{\bs k}]_x +[1-\hat{\bs p}\cdot\hat{\bs k}]\s_x+[\hat p_x \hat{\bs k}+ \hat{\bs p}\hat{k}_x]\cdot \bs \s \right\}$ \\
	    \hline
	    	7    & + & - & + & $i\g_1 \g_3 { = \tau_0 s_2}$ & $Q_7(\hat{\bs p},\hat{\bs k}) = {1\over 2}\left\{ -i [\hat{\bs p}\times \hat{\bs k}]_y +[1-\hat{\bs p}\cdot\hat{\bs k}]\s_y+[\hat p_y \hat{\bs k}+ \hat{\bs p}\hat{k}_y]\cdot \bs \s \right\}$ \\
	    \hline
	    	8    & + & - & + & $i\g_2 \g_1 { = \tau_0 s_3}$ & $Q_8(\hat{\bs p},\hat{\bs k}) = {1\over 2}\left\{ -i [\hat{\bs p}\times \hat{\bs k}]_z +[1-\hat{\bs p}\cdot\hat{\bs k}]\s_z+[\hat p_z \hat{\bs k}+ \hat{\bs p}\hat{k}_z]\cdot \bs \s \right\}$ \\
	    \hline
	    
	   9    & - & - & + &$i\g_0\g_1=-\tau_3 s_1$ & $Q_9(\hat{\bs p},\hat{\bs k}) ={\z\over 2}\left\{ (\hat {\bs p} +\hat {\bs k})_x-i [(\hat {\bs p} -\hat {\bs k})\times \bs \s]_x \right\}$ \\
		\hline
	   10    & - & - & + & $i\g_0\g_2=-\tau_3 s_2$ & $Q_{10}(\hat{\bs p},\hat{\bs k}) ={\z\over 2}\left\{ (\hat {\bs p} +\hat {\bs k})_y-i [(\hat {\bs p} -\hat {\bs k})\times \bs \s]_y \right\}$  \\ 
		\hline
		11    & - & - & + & $i\g_0\g_3= -\tau_3 s_3  $ & $Q_{11}(\hat{\bs p},\hat{\bs k}) = {\z\over 2}\left\{ (\hat {\bs p} +\hat {\bs k})_z-i [(\hat {\bs p} -\hat {\bs k})\times \bs \s]_z \right\}$ \\
		\hline
		12    & - & - & - &$i\g_0\g_5 = -\tau_2 s_0  $ & $Q_{12}(\hat{\bs p},\hat{\bs k}) =-i{\z \over 2}\left(\hat {\bs p}-\hat{\bs k} \right)\cdot \bs \s $ \\
		\hline
		13    & + & - & - &$i\g_1\g_5 = \tau_1 s_1 $ & $Q_{13}(\hat{\bs p},\hat{\bs k}) ={1\over 2}\left\{ i (\hat{\bs p}\times \hat{\bs k})_x-(\hat p_x \hat{\bs k}+\hat k_x \hat{\bs p})\cdot \bs \s +(1+\hat{\bs p}\cdot \hat{\bs k})\s_x\right\} $ \\
		\hline
		14    & + & - & - &$i\g_2 \g_5 = \tau_1 s_2  $ & $Q_{14}(\hat{\bs p},\hat{\bs k}) ={1\over 2}\left\{ i (\hat{\bs p}\times \hat{\bs k})_y-(\hat p_y \hat{\bs k}+\hat k_y \hat{\bs p})\cdot \bs \s + (1+\hat{\bs p}\cdot \hat{\bs k})\s_y\right\} $ \\
		\hline
	    15    & + & - & - &$i\g_3 \g_5 = \tau_1 s_3$ & $Q_{15}(\hat{\bs p},\hat{\bs k}) ={1\over 2}\left\{ i(\hat{\bs p}\times \hat{\bs k})_z-(\hat p_z \hat{\bs k}+\hat k_z \hat{\bs p})\cdot \bs \s +(1+\hat{\bs p}\cdot \hat{\bs k})\s_z\right\} $ \\
	    \hline
	\end{tabular} 
	\caption{ Table of the impurity scattering matrices in the orbital and band (MCBB) bases appearing in Eqs.~\eqref{eq:Disorder} and~\eqref{eq:Disorder_band_basis}. The table also lists the discrete symmetry properties of each scattering process under $\mc I = \g_0 = \tau_1 s_0$, $\mc T = \mc K \g_1 \g_3 = \mc K\tau_0 (-i s_2)$, and $\mc C = \gamma_5 = -\tau_3 s_0$, corresponding to inversion, time-reversal and chiral symmetries, respectively. $\zeta=+1/-1$ corresponds to Fermi level residing in   conduction/valence band, accordingly.}
	\label{tbl:Q}
\end{table*}

Before proceeding, we make a few important remarks regarding the choice of disorder potential in Eq.~\eqref{eq:Disorder}. First, we assume that the disorder is Gaussian correlated with zero mean, $\langle V_{l,m}\rangle = 0$. Next, we assume that there are no spatial correlations, and that different types of disorder do not correlate. The latter assumption implies that disorder potential does not break any symmetry {\it on average}, leading to  $\langle V_{l,m} V_{l',m'}\rangle = V_{m}^2 \d_{mm'} \d_{ll'}$. There is one exception, however, which requires clarification. In the absence of chiral symmetry, the density disorder ($m=0$) and the mass disorder ($m=1$) can, in principle, mix, since they belong to the same trivial representation. We note that this is the reason why mass disorder is always present in Bi$_2$Se$_3$, even if it respects time-reversal and inversion symmetries (as opposed to the claim made in Ref.~\cite{andersen2020generalized}). As we show in Sec.~\ref{sec:Tc},  however, the correlations between $m=0$ and $m=1$ disorder channels do not affect the scattering rate or superconductivity.

Second, a central assumption of our theory is that disorder naturally appears in the {\it orbital basis}. Indeed, the set of matrices $Q_m(\hat {\bs p}, \hat {\bs k})$ introduced in Eq.~(\ref{eq:Disorder_band_basis}) and listed in Table~\ref{tbl:Q} is the result of starting from the orbital basis and projecting disorder potential onto the MCBB on the Fermi surface. The additional momentum-dependent form-factors in the scattering matrices could have been easily overlooked if we started directly from the band basis, constructing a phenomenological picture of disorder~\cite{MineevSamokhinbook}. To emphasize this fact, we point out that even the density channel obtains non-trivial momentum dependence, which, as we  show below, plays a crucial role in protecting some unconventional pairing states from density disorder.

The last ingredient required to estimate the superconducting transition temperature is the attractive interaction which leads to the instability. We study the superconducting instability in the {\it band basis}, in the spirit of the Bardeen-Cooper-Schrieffer (BCS) theory.

We then decompose the interaction into the irreducible representations in the Cooper channel:
\be \label{eq:SI}
 \mc S_I = -{1\over 2}\sum_{\bs k ,\bs p,J} g_J \left[c^\dag _{\bs p}F_{J}^\dag(\hat {\bs p})   c^\dag_{-\bs p}\right] \left[c^{\vphantom\dagger}_{-\bs k}  F_{J}(\hat {\bs k}) c_{\bs k}\right],
\ee
where $F_J(\hat {\bs k})$ are form-factors in the MCBB corresponding to different representations $J$ of the relevant symmetry group, which are specfied in Table~\ref{Tbl:fmk}. A superconducting instability can occur in any one of the channels depending on the attractive strength of coefficients $g_J$. We are mainly interested in systems with large spin-orbit coupling, characteristic for topological materials, which do not have spin-rotational symmetry. That is why in this paper we focus on the fully isotropic $O(3)$ group of {\it joint} rotations of spin and momentum. Different representations are labeled by the total angular momentum $J$. (Note that within our notations $J$ labels both different representations and different components within the same representation.)


\begin{center}
\begin{table}[h!] 
	\begin{tabular}{ |c|c|c|c|l| }
		\hline
		$L$&$S$& $J$& $\mc I$ &Basis function\\
		\hline
		$0$&$0$&$0$&+&$F_{0g}=\sqrt{\dfrac{1}{2}} (-i\sigma^{y})$ \\
		1&1&0&-&$F_{0u}=\sqrt{\dfrac{1}{2}} (-i\sigma^{y} [{\hat{\bs{k}}}\cdot \bs{\sigma}])$ \\ 
		\hline
		&&&&$F_{11}=\sqrt{\dfrac{3}{4}} (-i\sigma^{y} [-\hat{k}_{z}\sigma^{y}+ \hat{k}_{y}\sigma^{z}])$\\ 
		$1$&$1$&$1$&-&$F_{12}=\sqrt{\dfrac{3}{4}} (-i\sigma^{y} [\hat{k}_{z}\sigma^{x}- \hat{k}_{x}\sigma^{z}])$\\
		&&&&$F_{13}=\sqrt{\dfrac{3}{4}} (-i\sigma^{y} [-\hat{k}_{y}\sigma^{x}+ \hat{k}_{x}\sigma^{y}])$\\
			\hline
		&&&& $F_{21}=\sqrt{\dfrac{3}{4}} (-i\sigma^{y} [\hat{k}_{x}\sigma^{y}+ \hat{k}_{y}\sigma^{x}])$\\
	
		&&&&$F_{22}=\sqrt{\dfrac{3}{4}} (-i\sigma^{y} [\hat{k}_{y}\sigma^{z}+ \hat{k}_{z}\sigma^{y}])$\\
		$1$&$1$&$2$&-&$F_{23}=\sqrt{\dfrac{3}{4}} (-i\sigma^{y} [\hat{k}_{x}\sigma^{z}+ \hat{k}_{z}\sigma^{x}])$\\
		&&&&$F_{24}=\sqrt{\dfrac{3}{4}} (-i\sigma^{y} [\hat{k}_{x}\sigma^{x}- \hat{k}_{y}\sigma^{y}])$\\
		&&&&$F_{25}=\sqrt{\dfrac{1}{4}} (-i\sigma^{y} [-\hat{k}_{x}\sigma^{x} -\hat{k}_{y}\sigma^{y}+2\hat{k}_{z}\sigma^{z}])$\\
		\hline
	\end{tabular} 
	\caption{Different representations of the time-reversal-invariant order parameters $F_{J}$. The labels $L$, $S$, and $J$ correspond to orbital, spin, and total angular momentum, respectively. Note that for every total angular momentum $J$ there are $2J+1$ states $|J,J_z \rangle$, where $J_z$ gets integer values between $-J$ and $J$. Thus, the two different states with $J=0$ both have $J_z=0$ and are distinguished by their transformation properties under inversion, either even~($g$) or odd~($u$). }
	\label{Tbl:fmk}
\end{table} 
\end{center}

\section{Computation of the scattering rate}\label{sec:Tc}
\begin{figure}
	\includegraphics[width=0.5\textwidth]{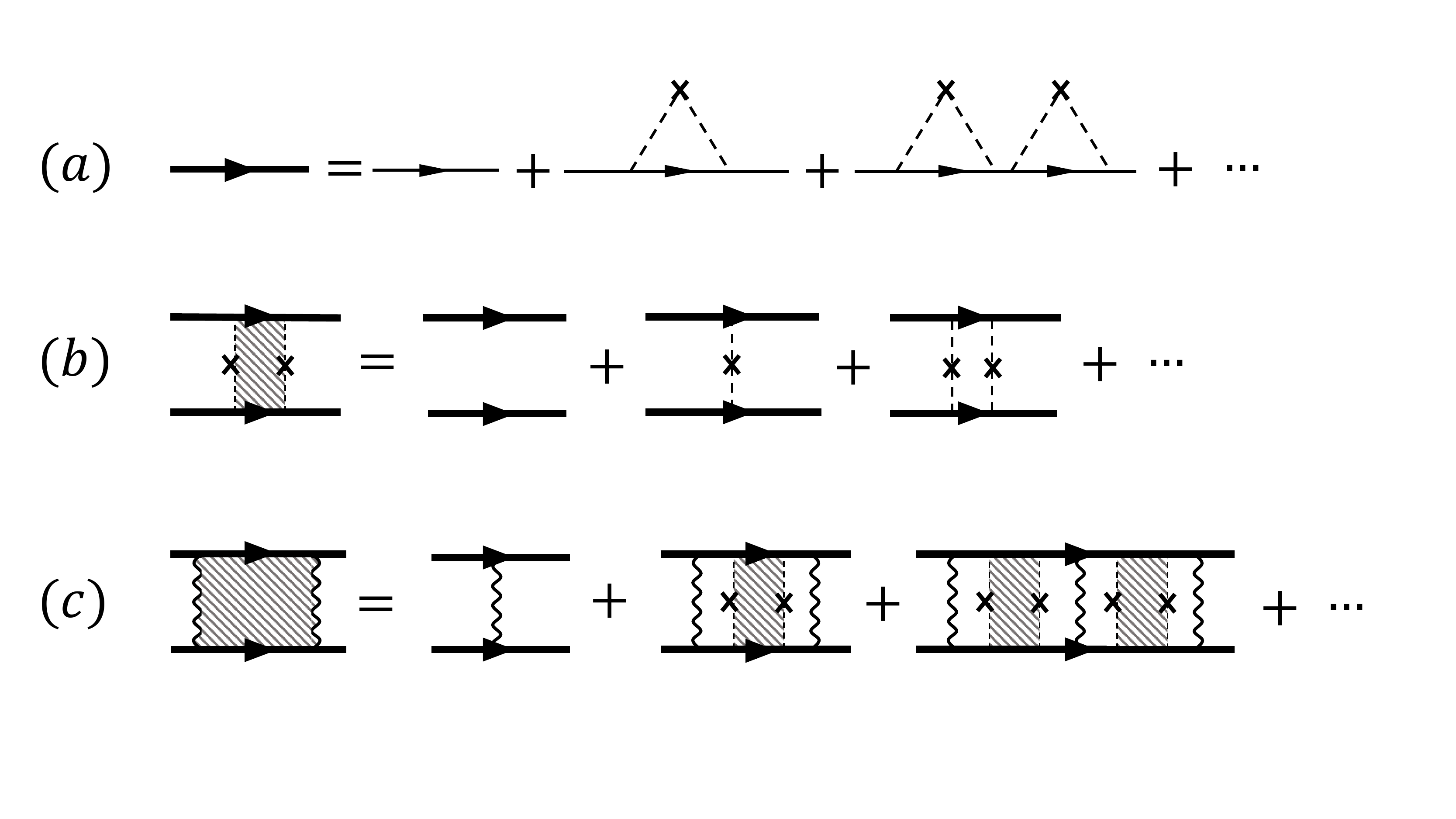}
	\caption{A diagrammatic representation of the summation of the Gor'kov ladder. (a) The summation over the scattering processes from all types of impurities within the first Born approximation, which leads to the self-energy correction to the full Green's function, Eq.~\eqref{eq:SelfEnergy1}.
     (b) The Cooperon  vertex correction $\mc B_J(i\w)$, Eq.~\eqref{eq:BC2}, which results from the summation over the bare pairing propagators $A(i\omega)$, Eq.~\eqref{eq:sus}. (c) The summation of the Gor'kov ladder. The building block of the ladder, $\mc P_J$, is given by the sum of $\mc B_J(i\w)$ over Matsubara frequencies, see Eq.~\eqref{eq:instability}. }
	\label{fig:diagrams}
\end{figure}

We now turn to the computation of the pair scattering rate $\G$, which enters the Abrikosov-Gor'kov theory and dictates the thermodynamics of superconductors. The procedure we employ consists of three main steps, which are described diagrammatically in  Fig.~\ref{fig:diagrams}. 

{\it Single-particle lifetime --} 
 We start with computing the single-particle lifetime. The bare electronic Green's function in the {\it band basis}  is given by ($\z=1$)
\begin{align}\label{eq:ProjectionGF}
G_0(i\w,\bs k)  =   {1\over i \w - v k+\e_F} .
\end{align} 
Summation of the diagrams in Fig.~\ref{fig:diagrams} (a) leads to a self-energy correction to the Green's function, 
$
G^{-1}(i\w,\bs k) = G_0^{-1}(i\w,\bs k) - \Sigma(i\w).
$
 Using Eqs.~\eqref{eq:Disorder_band_basis} and \eqref{eq:ProjectionGF}, we find that the self-energy is given by $\Sigma(i\w) = \sum_m \Sigma_m(i\w)$, with
\begin{align}\label{eq:SelfEnergy1}
\Sigma_{m}(i\w) 
&= {n_m V_m^2 \over 8\pi^3} \int d^3 p\, Q_{m}(\hat{\bs k},\hat{\bs p}) G_0(i\w,\bs p)Q_{m}(\hat{\bs p},\hat{\bs k})\nn\\
&= 
{n_m V_m ^2 k_F^2 \over 4\pi^2}\int_{-\infty} ^\infty {d p \over i\w-v p }  = - {i\,\mrm{sign \,(\w)} \over2\tau_m},
\end{align}
where 
\be \label{eq:tau}
\tau_m \equiv \frac1{\pi \nu_0 n_m V_{m}^2}.
\ee
Here $n_m = N_m/L^3$ is the density of impurities in channel $m$, { which arises after averaging over the positions of the impurities}, and $\nu_{0} = k_F^2 / 2\pi^2 v$ is the density of states at the Fermi level per pseudospin. We note a factor of $2$ difference in the definition of the scattering time compared to a parabolic band (see Appendix \ref{app:mass}), which is a feature of topological touching points of two bands.
Thus, we have obtained that the single-particle scattering rate decomposes into a sum over the different scattering channels:
\be\label{eq:Totaltau}
{1\over \tau} \equiv \sum_{m}{1\over \tau_m}.
\ee 

We recall that disorder is uncorrelated among different channels, which results from the assumption that disorder does not break any symmetry on average. Mass and density disorder are an exception, since they both belong to the trivial representation. {  However, we note that even if cross correlations between mass and density are present, for $\d = 0$ the cross term in Eq.~\eqref{eq:SelfEnergy1} vanishes. To see this we write the product 
\begin{align}
   \int d\W_{\bs p} Q^{\alpha \beta}_0(\hat{\bs k},\hat{\bs p})Q^{\beta \gamma}_1(\hat{\bs k},\hat{\bs p}) &= \int d\W_{\bs p}\langle \hat{\bs k}\alpha |\md 1| \hat{\bs p}\beta \rangle\langle \hat{\bs p}\beta |\g_0 | \hat{\bs k}\gamma \rangle\nn\\
   &= 2\pi \langle \hat{\bs k}\alpha |\g_0 | \hat{\bs k}\gamma \rangle = 0\nn,
\end{align} 
where we have used the identities $d\W_{\bs k} = d(\cos \t_{\bs k})d\phi_{\bs k}$, $\langle \hat{\bs k}\alpha |\g_0 | \hat{\bs k}\gamma \rangle = 0$, and $\int d\W_{\bs p}| \hat{\bs p}\beta \rangle\langle \hat{\bs p}\beta | = 2\pi \md 1$, and the summation over repeated index $\beta$ is implied.

}

{\it Vertex correction --} 
In addition to the single-particle processes, it is also important to take into account the effect of pair scattering. 
Namely, now we calculate the correction to the BCS vertex due to intermediate scattering on disorder. 
In the limit of weak disorder, $\e_F \tau \gg 1$, the most important correction to the Gor'kov ladder comes from the diagrams with non-intersecting impurity lines  (so-called Cooperon), as shown in Fig.~\ref{fig:diagrams} (b)~\cite{Gorkov2008}. 

To compute the disorder contribution to the vertex we need two ingredients. First, we calculate the bare propagator of a Cooper pair:
\begin{align}\label{eq:sus}
\begin{split}
A(i\w) = \sum_{\bs p} \mrm{Tr}\left[ G(i\w,\bs p)F_{J}^\dag(\hat{\bs p})  G^{\mrm T} (-i\w ,-\bs p)F_{J}(\hat{\bs p})\right] = \\ {k_F^2\over 2\pi ^2}\int_{-\infty}^\infty {dp\over \left(\w+{\mrm{sign\,(\w)}/2\tau}\right)^2+v^2 p^2}
= {2\pi\tau\nu_0\over1+2\tau |\w| }\,.
\end{split}
\end{align}
This propagator links between the scattering events on the Gor'kov ladder and corresponds diagrammatically to the first term on the r.h.s. of Fig.~\ref{fig:diagrams} (b) [i.e., it forms the legs of the ladder]. 
When decomposing a generic pairing interaction into the irreducible representations [as in Eq.~\eqref{eq:SI}], the Gor'kov ladder decomposes into scattering channels  of the orthogonal basis functions $F_J(\hat{\bs k})$, which are labeled by $J$. Thus, when writing Eq.~\eqref{eq:sus}, we assume that the Cooper pair propagator is contracted on both sides with the interaction lines in the corresponding  channel $J$. 

\begin{figure}
	\includegraphics[width=0.5\textwidth]{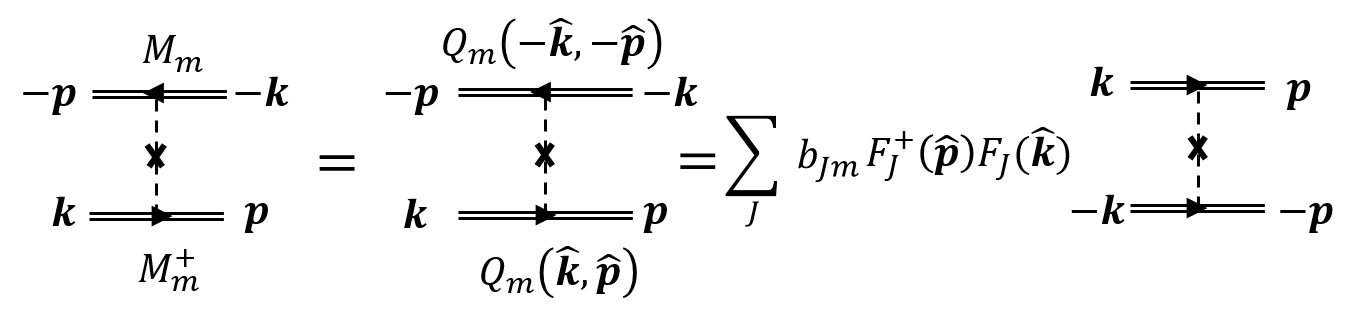}
	\caption{{ A diagrammatic representation of the decomposition from particle-hole to particle-particle channels, see Eq.~(\ref{eq:amp}). Matrices $M,$ $Q,$ $F,$  and $b$ are defined in Eqs.~\eqref{eq:Disorder}, \eqref{eq:Disorder_band_basis}, \eqref{eq:SI}, and~\eqref{eq:b} [see also Tables~\ref{tbl:Q}, \ref{Tbl:fmk}, and Eq.~(\ref{AppEq:bJm})]. }}
	\label{fig:flip}
\end{figure}

The next important ingredient  for calculating the Gor'kov ladder is the scattering amplitude from a single impurity in the particle-particle basis, which is the building element of a Cooperon and shown diagrammatically as the second term on the r.h.s. of Fig.~\ref{fig:diagrams} (b). Thus, we need the scattering amplitude of a Cooper pair with {\it any} momenta $\bs k$ and $-\bs k$ into a pair with any other momenta $\bs p$ and $-\bs p$ due to an impurity of type $m$. This amplitude is given by the product of two single-particle events:
\begin{align}\label{eq:amp}
n_m V_m^2\,Q_{m}^{\a\b}&(\hat{\bs p},\hat{\bs k}) Q_{m}^{\g\d}(-\hat{\bs p},-\hat{\bs k}) c^\dag_{\bs p \a}c_{\bs k \b}c^\dag_{-\bs p \g}c_{-\bs k \d} =  \\
&={1\over \pi \nu_0}\sum_{J} {b_{Jm}\over \tau_m} c^\dag_{\bs p} F^\dag_{J}(\hat {\bs p})c^\dag_{-\bs p}\,c_{-\bs k} F_{J} (\hat {\bs k})c_{\bs k},\nn
\end{align}
where 
\begin{align}\label{eq:b}
b_{Jm} = \int \frac{d\W_{\bs k} d\W_{\bs p}}{(4\pi)^2}\mrm{Tr}\left[Q_{m}(\hat{\bs{p}},\hat{\bs{k}})F_{J}^\dag(\hat{\bs{k}})Q_{m}^{\mrm T}(-\hat{\bs{p}},-\hat{\bs{k}})F_{J}(\hat{\bs{p}}) \right]
\end{align}
is a matrix of weights corresponding to the conversion from the particle-hole to particle-particle basis (similar to the Fierz identity~\cite{savary2017superconductivity}) and $d\W_{\bs k} = d(\cos \t_{\bs k})d\phi_{\bs k}$ is the solid angle element. The matrix in Eq.~\eqref{eq:b} is given explicitly in Appendix \ref{app:reps}. Here we only note that $|b_{Jm}|\leq 1/2$. {The procedure of decomposition from particle-hole to particle-particle basis is shown schematically in Fig.~\ref{fig:flip}.}

Contracting the two ingredients, Eq.~\eqref{eq:sus} and Eq.~\eqref{eq:amp}, and using the orthogonality of the superconducting form-factors $F_J(\hat{\bs k})$, we obtain for the disorder corrected block of the Gor'kov ladder, shown schematically in Fig.~\ref{fig:diagrams} (b): 
\be \label{eq:BC2}
\mc B_J(i\w) = {A(i\w) \over 1-{ A(i\w)}\sum_{m}{b_{Jm}/ {  \pi \nu_0 \tau_{m}}}  }={\pi \nu_0 \over {\G_J/2} + |\w|},
\ee
where
\be\label{eq:Gamma}
\G_J =\sum_{m}\G_{Jm}\;\; ;\;\; \G_{Jm}\equiv {1-2b_{Jm}\over\tau_m}
\ee
is the pair scattering rate, which is a sum of independent scattering rates $\G_{Jm}$ originating from the different intra- and interorbital disorder channels $m$.
The values for the partial pair scattering rates from  Eq.~\eqref{eq:Gamma} are { the main result of this paper} and are listed in Table~\ref{Tbl:rhoc}. 

\begin{figure}[h]
    \centering
    \includegraphics[width=0.5\textwidth]{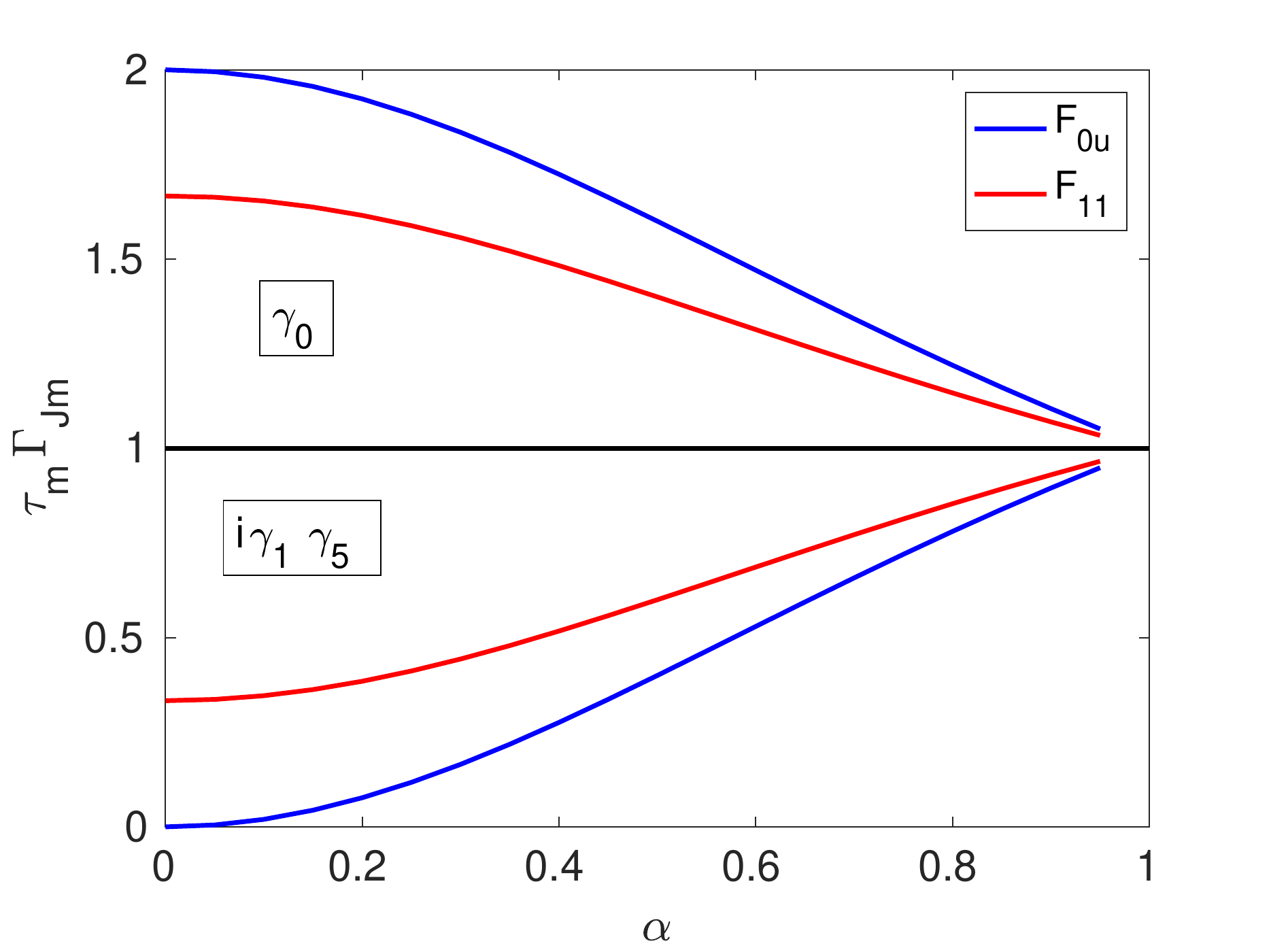}
    \caption{The dependence of the pair scattering rate $\tau_m\Gamma_{Jm}$ on Dirac mass $\a = \d/\e_F$ (where $\e_F = \sqrt{(vk_F)^2 + \d^2}$), for odd-parity pairings $F_{0u}$ (blue) and $F_{11}$ (red). The upper half is the effect of mass disorder $\gamma_{0}$, whereas the lower half shows magnetic disorder $i\gamma_{1}\gamma_{5}$. While the former has the most severe effect on $F_{0u}$, this pairing channel is protected against the latter due to the $\mathcal{C}\mathcal{T}$ symmetry. Nevertheless, they all approach the Larkin's result $\tau_m\Gamma_{Jm}=1$ (black) as the mass goes to infinity ($\alpha=1$).} 
    \label{fig:GammaMass}
\end{figure}

{ As mentioned below Eq.~\eqref{eq:S0}, up to this point we have focused on the case of zero mass, $\delta = 0$. In this case the product of  $\tau_m\G_{Jm}$ takes the universal rational values presented in Table~\ref{Tbl:rhoc}. To discuss how these numbers vary with a finite mass $\delta$ we introduce a parameter $\alpha = \delta/\epsilon_{F}$, which ranges between $0$ (no mass) and $1$ (infinite mass).
As a consequence, the single particle rates in Eq.~\eqref{eq:tau} and the pair breaking rates in Eq.~\eqref{eq:Gamma} are modified to Eq.~\eqref{eq:MassTau} and Eq.~\eqref{Eq:Gamma*}, respectively (for details, see Appendix~\ref{app:mass}). We can distinguish three cases. The first one is the case of $s$-wave pairing ($F_{0g}$), in which the mass does not affect the results in Table~\ref{Tbl:rhoc}. The second case is $J \ne 0g$ (non-$s$-wave) and the disorder matrices are inversion symmetric (i.e., $\mc I^{-1}M_m \mc I = M_m$). In this case, the values of $\G_{Jm}\tau_m$ continuously interpolate between those in Table~\ref{Tbl:rhoc} and the asymptotic value without spin-orbit coupling $\G_{Jm}\tau_m = 1$ which was computed by Larkin~\cite{Larkin1965}. The functional form of this interpolation is given by Eq.~\eqref{Eq:Gamma*} and plotted in Fig.~\ref{fig:GammaMass}. Finally, the third case is $J \ne 0g$ and disorder matrices  are odd under inversion (i.e., $\mc I^{-1}M_m \mc I = -M_m$). In this case, the matrices $M_m$ act as purely interband operators within the Bloch wave functions when taking the limit $\a\to1$. Consequently, $\G_{Jm}$ and $1/\tau_m$ go to zero as the conduction and valence bands become infinitely separated, i.e., when $\alpha$ increases. It is interesting, however, that their ratio remains the same for all $\a$ and is given in Table~\ref{Tbl:rhoc}.}

\begin{center}
	\begin{table*}
		\begin{tabular}{ |c|c|c|c|c|c|c||c|c|c||c|c|c||c|c|c||c||c|c|c| }
			\hline
			& & & &$\md 1$ & $\gamma_{0}$&$ \gamma_{5}$& $\gamma_{1}$&$\gamma_{2}$&$\gamma_{3}$ &$i\g_3\g_2$&$i\g_1\g_3$&$i\g_2\g_1$&$i\g_0\g_1$&$i\g_0\g_2$&$i\g_0\g_3$&$i\g_0\g_5$&$i\g_1\g_5$&$i\g_2\g_5$&$i\g_3\g_5$  \\\hline
			&&&$m$& $0$ & $1$  & $2$ & $3$& $4$& $5$&$6$&$7$&$8$&$9$& $10$&
			$11$&$12$&$13$&$14$&$15$ \\
			\hline
			 &&&$\mc T$& + & +  & + & +& +& +&-&-&-&-&- &- &-&-&-&- \\
			
			\hline 
			&&&  $\mc I$& + & +  & - & -& -& -&+&+&+&-&- &- &-&+&+&+ \\
			
			\hline 
		&&&	$\mc C$& + & -  & + & -& -& -&+&+&+&+&+ &+ &-&-&-&-\\
			\hline
			\hline 
			$L$&$S$&$J$& &  &   &  & & & &&&&& & &&&& \\
			
			\hline 
			0 & 0 & $0g$&$F_{0g}$& $0$& $0$ & $0$ & $0$&$0$&$0$ & \cellcolor{p2}{$2$}& \cellcolor{p2}{$2$} & \cellcolor{p2}{$2$} &\cellcolor{p2}$2$&\cellcolor{p2}$2$&\cellcolor{p2}$2$&\cellcolor{p2}$2$&\cellcolor{p2}$2$&\cellcolor{p2}$2$&\cellcolor{p2}$2$\\
			1 &1 & $0u$&$F_{0u}$& $0$&\cellcolor{p2}{$2$} &$0$ & \cellcolor{p2}{$2$} &\cellcolor{p2}{$2$}& \cellcolor{p2}{$2$}&\cellcolor{p2}{$2$}&\cellcolor{p2}{$2$}&\cellcolor{p2}{$2$}&\cellcolor{p2}$2$&\cellcolor{p2}$2$&\cellcolor{p2}$2$&$0$&$0$&$0$&$0$\\
		 1& 1&$11$ &	$F_{11}$ &\cellcolor{p13}{$1/3$}&\cellcolor{p53}{$5/3$}&\cellcolor{p53}{$5/3$}& \cellcolor{p13}{$1/3$}&\cellcolor{p53}{$5/3$} &\cellcolor{p53}{$5/3$}&\cellcolor{p53}{$5/3$}&\cellcolor{p13}{$1/3$}&\cellcolor{p13}{$1/3$}&\cellcolor{p13}$1/3$&\cellcolor{p53}$5/3$&\cellcolor{p53}$5/3$&\cellcolor{p53}$5/3$&\cellcolor{p13}$1/3$&\cellcolor{p53}$5/3$&\cellcolor{p53}$5/3$\\
			1 & 1&$12$ &$F_{12}$ &\cellcolor{p13}{$1/3$}&\cellcolor{p53}{$5/3$}&\cellcolor{p53}{$5/3$}&  \cellcolor{p53}{$5/3$}&\cellcolor{p13}{$1/3$} &\cellcolor{p53}{$5/3$}&\cellcolor{p13}{$1/3$}&\cellcolor{p53}{$5/3$}&\cellcolor{p13}{$1/3$}&\cellcolor{p53}$5/3$&\cellcolor{p13}$1/3$&\cellcolor{p53}$5/3$&\cellcolor{p53}$5/3$&\cellcolor{p53}$5/3$&\cellcolor{p13}$1/3$&\cellcolor{p53}$5/3$\\
			1 & 1&$13$ &$F_{13}$ &\cellcolor{p13}{$1/3$}&\cellcolor{p53}{$5/3$}&\cellcolor{p53}{$5/3$}&  \cellcolor{p53}$5/3$&\cellcolor{p53}$5/3$ &\cellcolor{p13}{$1/3$}&\cellcolor{p13}{$1/3$}&\cellcolor{p13}{$1/3$}&\cellcolor{p53}{$5/3$}&\cellcolor{p53}$5/3$&\cellcolor{p53}$5/3$&\cellcolor{p13}$1/3$&\cellcolor{p53}$5/3$&\cellcolor{p53}$5/3$&\cellcolor{p53}$5/3$&\cellcolor{p13}$1/3$\\
			1&1&$2$&$F_{2}$&\cellcolor{p1}$1$&\cellcolor{p1}$1$&\cellcolor{p1}$1$&\cellcolor{p1}$1$&\cellcolor{p1}$1$&\cellcolor{p1}$1$&\cellcolor{p1}$1$&\cellcolor{p1}$1$&\cellcolor{p1}$1$&\cellcolor{p1}$1$&\cellcolor{p1}$1$&\cellcolor{p1}$1$&\cellcolor{p1}$1$&\cellcolor{p1}$1$&\cellcolor{p1}$1$&\cellcolor{p1}$1$\\
			\hline
		\end{tabular}
		\caption{Top: Indicates different types of intra- and interorbital scattering matrices and their properties under the discrete symmetries. The scattering matrices are labeled by $m=0,\ldots,15$ and appear as gamma matrices and their products. $\mc T = \mc K \g_1\g_3$, $\mc I=\g_0$, and $\mc C=\g_5$ correspond to time-reversal, inversion, and chiral symmetries, respectively. 
		Bottom: The values of the dimensionless pair scattering rate $\tau_{m}\G_{Jm} = 1-2b_{Jm}$ [where the matrix $b_{Jm}$ is defined in Eq.~\eqref{eq:b}] for different superconducting pairing states. The labels $L$, $S$, and $J$ correspond to orbital, spin, and total angular momentum, respectively (we use same convention as in Ref.~\cite{fu2015parity}). It is clear that the $F_{0g}$ state is protected from disorder that respects $\mc T$ symmetry (Anderson's theorem), while the $F_{0u}$ state is protected from disorder that respects $\mc{CT}$ symmetry. In the last line, $F_2$ implies all the pairing states with $J=2$, i.e., $F_{21}-F_{25}$. The result for all of these states is the same.}
		\label{Tbl:rhoc}
	\end{table*}
	
\end{center}

\begin{figure}[h]
	\includegraphics[width=0.5\textwidth]{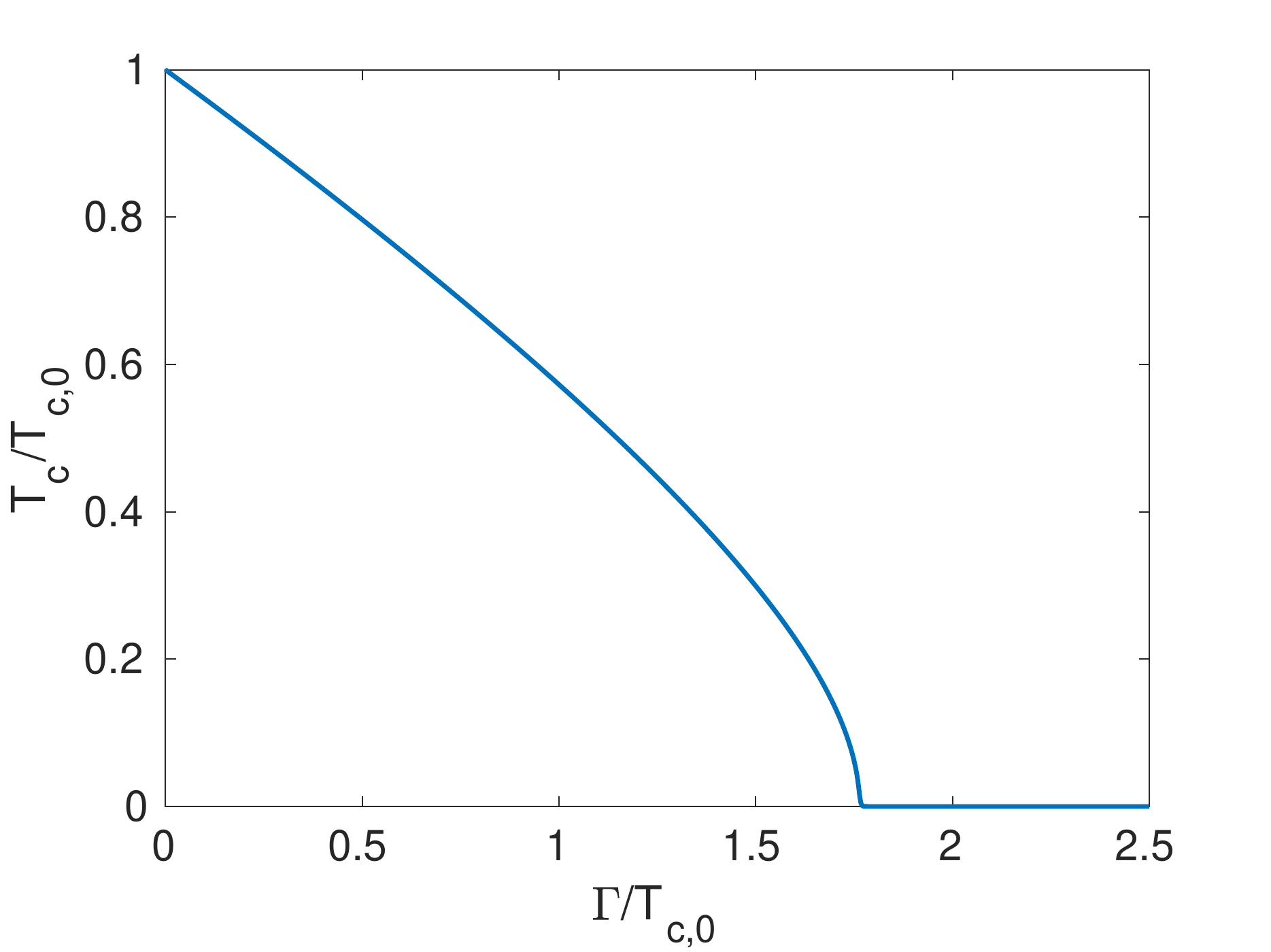}
	\caption{$T_c/T_{c, 0}$ as a function of $\Gamma/T_{c,0}$ obtained from the solution of Eq.~\eqref{eq:criticalT}. }
	\label{fig:Tc1}
\end{figure}

{\it Computation of $T_c$ --} 
{ The  final step of the calculation is to use the disorder-modified interaction in the superconducting channel $J$ to compute the renormalized pairing vertex. To do that, we insert a Cooperon in each block of the Gor'kov ladder [this step gives us factor $\mc B_J(i\w)$], perform the summation over intermediate Matsubara frequencies $\omega_n = \pi T (2n+1)$, and sum up all the blocks [as shown in Fig.~\ref{fig:diagrams} (c)]. The result reads as

\be \label{eq:instability}
    \tilde g_J =  {g_J/( 1+ g_J \mc P_J)},
\ee
where 
$ \mc P_J = -T \sum_{\omega_n} \mc B_{J}(i\omega_n)$, and $T$ is temperature. The transition temperature $T_{c, J}$ in the channel $J$ is determined as a singularity (vanishing denominator) in Eq.~\eqref{eq:instability}, leading to the result that has the Abrikosov-Gor'kov form  

\begin{align}\label{eq:criticalT}
\log \left(\dfrac{T_{c,J}}{T_{c,J,0}} \right) = \Psi(1/2)-\Psi\left(1/2+{\Gamma_J}/{4\pi T_{c,J}}\right),
\end{align}
where $\Psi(x)$ is the digamma function, and $T_{c,J,0}$ is the transition temperature in the absence of any disorder. For the detailed calculation of the transition temperature, see Appendix~\ref{app:Tc}.  An alternative derivation of this result via the method of Gor'kov Green's functions (which also gives a solution below $T_c$) is presented in Appendix~\ref{app:gap}.

The  solution of Eq.~(\ref{eq:criticalT})  predicts that superconductivity is completely suppressed at 
\be 
\Gamma_{J \text{cr.}} = \pi e^{-\gamma_e} T_{c,J,0} \approx 1.76 \, T_{c,J,0}, \label{Eq:Gammacritical}
\ee
($\gamma_e\approx 0.577...$ is the Euler's constant), as shown in Fig.~\ref{fig:Tc1}.
}

\section{Results} \label{sec:Results}

In Section~\ref{sec:Tc}, we described how the pair scattering rates in Eq.~\eqref{eq:Gamma} affect the transition temperature. As we show in Appendix~\ref{app:gap}, the effect of these rates is actually much more general as they dictate the entire low-temperature thermodynamics of these superconductors (up to phase fluctuation effects)~\cite{abrikosov1960contribution,Gorkov2008}. For example, we recall that the gap may close in the superconducting state when disorder is sufficiently strong.  

Having established the importance of the pair scattering rates in Eq.~\eqref{eq:Gamma} for superconductivity in doped Dirac systems, we now turn to discuss their value for different pairing states and different interorbital disorder potentials. The results are summarized in Table~\ref{Tbl:rhoc}, which  includes both non-magnetic ($m=0,\ldots,5$) and magnetic ($m = 6,\ldots,15$) impurities. 

As mentioned above, the elements of the matrix $b_{Jm}$ in Eq.~\eqref{eq:b} range between $1/2$ and $-1/2$ (see Appendix~\ref{app:reps}). Consequently, the rates $\G_{Jm}$  appearing in Table~\ref{Tbl:rhoc}, which shows the values of $1-2b_{Jm}$, range between $0$ and twice the single particle scattering rate $1/\tau_{m}$. The former implies that disorder in channel $m$ does not affect superconductivity in channel $J$, while the latter corresponds to the most severe effect possible. 

Indeed, for the $s$-wave channel ($J=0g$), we find that the pair scattering rate vanishes, $\G_{0g} = 0$, for all the $\mc T$-even disorder matrices  $m=0,\ldots, 5$,
which is manifestly the Anderson's theorem for non-magnetic impurities.

Additionally, we recover the well known Abrikosov-Gor'kov result for magnetic impurities, that the pair scattering rate is twice that of single particles~\cite{abrikosov1960contribution}, such that overall the pair-breaking rate is given by
\be\label{eq:Gamma_0g}
\G_{0g} = \sum_{m=6}^{15} {2\over \tau_m}\,.
\ee

Another limit of interest is the odd-parity state with total angular momentum zero ($J =0u$), which is equivalent to the B-phase of superfluid $^3$He~\cite{salomaa1986topological}. We notice that, similar to the s-wave state $F_{0g}$, this pairing state is completely protected from certain types disorder, namely $m=0,2$ and $12,\ldots,15$. Inspecting Table~\ref{Tbl:rhoc} we identify that the common symmetry of these disorder potentials is that they are all even under the product of chiral and time reversal $\mc{CT}$. What makes this result even more interesting is that some of the $\mc{CT}$-even matrices are $\mc T$-odd. Thus, the $F_{0u}$ is protected from certain types of magnetic impurities. We identify this protection with similar results for superconductors with multiple Fermi surfaces~\cite{golubov1997effect}.  

To understand this protection we now show that the $F_{0u}$ is essentially a singlet pairing state between partners related to each other by $\mc{CT}$ symmetry. Thus, in complete equivalence to the Anderson's original argument~\cite{anderson1959theory}, it follows that as long as the disorder potential does not violate $\mc{CT}$ symmetry, we can always pair $\mc{CT}$ partners in the basis that diagonalizes the disorder potential { (see Appendix~\ref{app:AndersonCT})}.

Let us show that the $F_{0u}$ pairing state is indeed a singlet state of $\mc{CT}$ partners. This is most easily seen in the orbital basis. We find it convenient to rotate the orbital basis by $\pi/2$ about the $\tau_2$ axis first. This transforms from the basis of chirality to the basis of parity (i. e., the orbitals are labeled by their parity $\tau = \pm$). Note that this does not affect the operation of time-reversal $\mc T$. Then the action of chiral symmetry is implemented by $\mc C = \tilde \g_5 = \tau_1s_0$ and the corresponding pairing state $F_{0u}$ is
\begin{widetext}
\begin{align}
\D_{0u}(\bs k) = {1\over 2}\psi_{\bs k}  \, \tau_1(-is_2)\,\psi_{-\bs k}  
={1\over 2}\left[\psi_{\da +}(\bs k)\psi_{\ua -}(-\bs k)-
\psi_{\ua -}(\bs k)\psi_{\da +}(-\bs k)+
\psi_{\da -}(\bs k)\psi_{\ua +}(-\bs k)-
\psi_{\ua +}(\bs k)\psi_{\da -}(-\bs k)
\right],
\end{align}
\end{widetext}
where $\psi_{s\tau}(\bs k)$ is a field operator in the rotated basis. Inspecting this pairing state, it is evident that it is fully antisymmetric and that each term consists of a pair of operators related to each other by $\mc{CT}$ symmetry.

However, the $F_{0u}$ state becomes vulnerable to disorder when $\mc{CT}$ symmetry is not present, such as in doped Bi$_2$Se$_3$. In that case mass belongs to the same representation as density and is always present.  Moreover, we argue that Dirac materials are often polar ionic crystals (e.g. Bi$_2$Se$_3$, SnTe, PdTe etc.), therefore, it is likely that the disorder potential also induces dipolar moments of type $m=3,4,5$. 
This argument should be contrasted with the claim made in Ref.~\cite{andersen2018nematic}, where it was stated that only density disorder should be present. 
Overall, we find that the pair-breaking rate in the $F_{0u}$ channel equals
\be
\G_{0u} ={2\over \tau_1}+\sum_{m=3}^{11}{2\over \tau_m}.
\ee

It should also be noted that the authors of Ref.~\cite{michaeli2012spin} were the first ones to identify that the $F_{0u}$ state can be protected from disorder in the massless limit.
However, they concluded that it is protected by $\mc C$ symmetry. As we show here, it is actually protected by $\mc{ CT}$ symmetry. To emphasize this distinction between the two, we point out that the $F_{0u}$ state is immune to some disorder potentials that are odd under $\mc C$, such as $m = 12,\ldots,15$.

Next, we consider  the odd-parity pairing states with total angular momentum $J=1$.
These states (more accurately, the nematic $E_u$ states of the $D_{3d}$ symmetry group of Bi$_2$Se$_3$, which derive from the $J=1$ representation by breaking the rotational symmetry from spherical to trigonal) are of special interest experimentally, since they are considered to be the pairing state in doped Bi$_2$Se$_3$~\cite{matano2016spin,yonezawa2017thermodynamic,WillaWillaetal2018,PandeVisseretal2016,NikitindeVisseretal2016,AsabaLuLietal2017,shen2017nematic,Taoetal2018,Funematic2014,VenderbosKoziiFu2016}. We find that all disorder channels affect superconductivity with this pairing symmetry, and the dimensionless rate $\tau_m \G_m$ takes two possible values $5/3$ and $1/3$.
Thus, depending on the relative weight in these channels, the scattering rate can vary significantly. 
In particular, for density disorder, the rate $\G_{1j,0} = 1/3\tau_0$ ($j=1-3$ here, see Tables~\ref{Tbl:fmk} and~\ref{Tbl:rhoc}) is much smaller than in systems without spin-orbit coupling, where it is expected to be $1/\tau_0$~\cite{Larkin1965,mackenzie1998extremely}. However, our results are not consistent with experiments that find this state to be protected~\cite{smylie2017robust,andersen2020generalized} and also not consistent with previous theoretical work where density disorder was considered~\cite{nagai2015robust,andersen2020generalized}. Even more so when considering the  $m>0$ channels which more harmful. For example, if polar disorder is present, then an average over all possible directions gives $\G_{1j,3-5} = {11/9 \tau_3}$, where we assume $\tau_3 = \tau_4 = \tau_5$ by symmetry, and, again, $j=1-3$.

\section{Discussion} \label{sec:Discussion}
In this paper, we construct a rigorous method to evaluate the effect of  both intra- and interorbital disorder on superconductivity in doped Dirac materials. We argue that generic disorder potential  always induces interorbital scattering processes given by Eq.~\eqref{eq:Disorder} and listed in Table~\ref{tbl:Q}. We compute the contribution of each type of intra- and interorbital scattering channels to the pair-breaking rate for a given pairing potential, which dictates the entire thermodynamics of a superconductor. This result is summarized in Table \ref{Tbl:rhoc}. 
Our main conclusions from this analysis are as follows:
\begin{itemize}
    \item{We have found a version of the Anderson's theorem which is based on the pairing of $\mc{CT}$-partners rather than $\mc T$-partners ($\mc{C}$ is chiral and $\mc{T}$ is time-reversal symmetry). The odd-parity state with total angular momentum zero ($J = 0u$) is such a pairing state. Consequently, it is protected from disorder that respects $\mc{CT}$ symmetry, which includes certain types of $\mc T$-odd impurities. This also generalizes the results of Ref.~\cite{michaeli2012spin}. It is interesting to understand in the future if such a symmetry can exist (or nearly exist) in a solid state material.  }
    \item{As expected from the Anderson's theorem, the $s$-wave ($J=0g$) pairing state is protected from all non-magnetic scattering processes, including interorbital ones. }
    \item{The $J=1$ states, which can be considered as the $O(3)$ analog of the multicomponent nematic candidate state for doped Bi$_2$Se$_3$ ~\cite{matano2016spin,yonezawa2017thermodynamic,WillaWillaetal2018,PandeVisseretal2016,NikitindeVisseretal2016,AsabaLuLietal2017,shen2017nematic,Taoetal2018,Funematic2014,VenderbosKoziiFu2016}, are not protected. We find that at most their pair breaking rate is suppressed by a factor of three compared to systems without spin-orbit coupling~\cite{Larkin1965}. This raises a question regarding the protection of this state observed in experiment~\cite{smylie2017robust,andersen2020generalized}.} 
\end{itemize}

We emphasize that the results presented in Table.~\ref{Tbl:rhoc} are based on a model where the mass term in the single-particle Hamiltonian, Eq.~\eqref{eq:S0}, was assumed to be zero. It should be noted, however, that in the majority of doped topological materials which become superconducting, such a mass term exists. 
{The dependence of these numbers on the mass is discussed in Appendix \ref{app:mass} and plotted in Fig.~\ref{fig:GammaMass}. For inversion-symmetric disorder potentials the inclusion of this term modifies these numbers towards their known values without spin-orbit coupling~\cite{abrikosov1960contribution,Larkin1965}. In particular, it removes the protection of the $F_{0u}$ state. For inversion-odd disorder, on the other hand, the values in Table~\ref{Tbl:rhoc} remain unchanged. For more details we refer the reader to Appendix~\ref{app:mass}.}

The analysis performed in this work assumes a short-ranged disorder potential. However, in doped materials, one may also anticipate a correlated potential emerging from charged impurities~\cite{skinner2014coulomb}. Therefore, it is important to also understand the influence of a soft potential on superconductivity. 

Another question which was not addressed in this paper is the 
microscopic origin of pairing and how it is affected by disorder~\cite{finkel1994suppression}. In particular, doped topological materials are characterized by small electronic density and small density of states. As a result, the pairing interaction must be more singular~\cite{kozii2019superconductivty}. Such an interaction is expected to be sensitive to the presence of disorder~\cite{lee1985disordered}. 

Finally, our results hold only for the case of a finite Fermi energy and weak disorder, implying $\e_F \gg \D$ and $\e_F \tau \gg 1$. It would be interesting to consider the limit of low density, where both conduction and valence bands are important. In this limit however, a number of issues arise. First, the assumption $k_F l \gg 1$ breaks down and therefore the self-energy and Born approximations are invalid and other approaches (e.g. the replica approach) must be used~\cite{nandkishore2013superconductivity,yerzhakov2018disorderd,boyack2020quantum}. Second, the omission of one of the bands is invalid and all four bands must be taken into account. Finally, we note that this scenario is however very exotic and, as discussed above, requires long-ranged interactions~\cite{kozii2019superconductivty}.

Looking forward, we argue that our theory is useful to many other systems with strong orbital hybridization. In particular, Eq.~\eqref{eq:b} is easily generalizable to different Hamiltonians and reduced dimensions.
Of special interest are semimetallic systems, including a quadratic band touching point relevant to the half-Heusler compounds~\cite{nakajima2015topological}, line-node semimetals, Weyl semimetals that emerge when inversion is broken in a Dirac material~\cite{kozii2019superconductivty}, and higher-order band touching points~\cite{bradlyn2017topological}.  

We also note that the results in Table~\ref{Tbl:rhoc} suggest that the effects of disorder on unconventional pairing states depend strongly on the microscopic nature of disorder (we note that a similar disorder dependent pair-breaking rate has been argued to exist in superfluid $^3$He in a nematic aerogel~\cite{fomin2018analog}). This opens an interesting avenue to manipulate the superconducting ground state by selectively inducing specific types of disorder potentials.

Before concluding this paper, we note that arguments for robustness of unconventional superconductivity to disorder were recently cast in terms of the so-called {\it superconducting fitness}~\cite{cavanagh2020robustness,andersen2020generalized,timmons2020electron}, which was first discussed in Ref.~\cite{ramires2018tailoring} in the context of clean systems.  An intuitive understanding of the fitness can actually be obtained based on Ref.~\cite{potter2011engineering}, where the affects of disorder on superconductivity were assessed by looking at the minimal excitation of a system and comparing it with the clean limit. As shown in Ref.~\cite{timmons2020electron} this translates to the condition that the Hamiltonian including disorder commutes with the gap function, $[\hat H + \hat V , \hat \D]_{\pm} = 0$, where the $\pm$ stands for commutation/anticommutation for TR even and TR odd disorder, respectively. Our results are consistent with this picture.

\section{Acknowledgments}
We are grateful to Rafael Fernandes, David M\"ockli, Maxim Khodas, Mathias Scheurer, David Cavanagh,  and Yuki Nagai for helpful discussions. J. R. and D. D. acknowledge the support of the Israeli Science Foundation under grant No. 967/19. V. K. was supported by the Quantum Materials program at LBNL, funded by the US Department of Energy under Contract No. DE-AC02-05CH11231. 


\appendix

\section{The effect of a finite mass}\label{app:mass}

In the main text  we considered the case of massless particles (i.e., $\delta=0$). As shown in Fig.~\ref{fig:GammaMass}, in the case of a finite mass the results in Table~\ref{Tbl:rhoc} will be continuously modified towards the well known results for systems without spin-orbit coupling (with the exception of inversion-odd disorder potentials).
For the sake of completeness we now compute the pair scattering rate $\G$ for massive particles. The eigenstates in the conduction band in
 the MCBB now have the form~\cite{kozii2019superconductivty} 
 
\begin{align}
&|\hat{\bs k},1,\a\rangle = {1\over 2}
\begin{pmatrix}
\beta_{+} - \beta_{-}\hat k_z, &
-\beta_{-}\hat k_+,&
\beta_{+}+\beta_{-}\hat k_z,&
\beta_{-}\hat k_+
\end{pmatrix}^T, \nonumber
\\
&|\hat{\bs k},2,\a\rangle = {1\over 2}
\begin{pmatrix}
-\beta_{-}\hat k_-, &
\beta_{+} + \beta_{-}\hat k_z,&
\beta_{-}\hat k_-,&
\beta_{+} - \beta_{-}\hat k_z
\end{pmatrix}^T,
\end{align}
where $\beta_{\pm} = \sqrt{1\pm \alpha}$ and we defined $\alpha\equiv \delta/\e_F$ with $\epsilon_{F} = \sqrt{\delta^2+(vk_{F})^2}$. 
The bare Green's function is given by 

\be  
G_0(i\omega, \bk) = \frac1{i\omega - \xi_\bk}, 
\ee
with $\xi_\bk = \sqrt{(vk)^2 + \delta^2} - \e_F$. The self-energy  then equals
\begin{align}
\Sigma_{m}(i\w) 
&= {n_m V_m^2 \over 8\pi^3} \int d^3 p\, Q_{m}(\hat{\bs k},\hat{\bs p}, \alpha) G_0(i\w,\bs p)Q_{m}(\hat{\bs p},\hat{\bs k},\alpha)\nn\\
&= 
{n_m V_m ^2 k_F^2 \over 4\pi^2 v_F }\eta_{m}\int_{-\infty} ^\infty {d \xi_p \over i\w- \xi_p},
\end{align}
where $v_F={v^2 k_{F}/{\e_{F}}}$ and $Q^{ij}_{m}(\hat{\bs k},\hat{\bs p},\a) = \langle \hat{ \bs{k}},i,\a| M_m |\hat{\bs p}, j, \a \rangle$. Consequently, we obtain
\begin{align}
    \Sigma_{m}(i\w)  = - {i\,\mrm{sign \,(\w)} \over2\tau^{\star}_m},
\end{align}
where 
\begin{align}\label{eq:MassTau}
    \tau^{*}_m \equiv \frac1{\eta_{m} \pi \nu^*_0 n_m V_{m}^2},
\end{align} 
and $\nu^*_0 = k_F^2/2\pi^2 v_F$.
The factor $\eta_m$ is given by
\begin{align}\label{eq:AppEta}
   \eta_{m} &= \dfrac{1}{4\pi} {\Tr} \left[\int d\Omega_{\textbf{p}}  Q_{m}(\hat{\bs k},\hat{\bs p}, \alpha) Q_{m}(\hat{\bs p},\hat{\bs k}, \alpha)\right] = \nonumber \\ &=  1 + I_m\alpha^2,
\end{align}
where $I_m = \pm 1$ is the inversion eigenvalue of the corresponding disorder matrix listed in Tables~\ref{tbl:Q} and~\ref{Tbl:rhoc} (i.e., $\mc I ^{-1} M_m \mc I = I_m M_m$). Note that trace adds an extra factor of 2 in the above expression, and we used the fact that the expression under the trace is proportional to the unity matrix.

The pairing form factors $F_J$ from Table~\ref{Tbl:fmk} are not changed by a finite mass and the whole procedure for calculating the effect of disorder is analogous to that for the massless case.  In particular, instead of Eq.~\eqref{eq:BC2} we find

\be 
\mc B_J(i\w) = {A(i\w) \over 1-{ A(i\w)}\sum_{m}{b^*_{Jm}/ {  \pi \nu^*_0 \tau^*_{m} \eta_m}}  }={\pi \nu_0^* \over {\G_J^*/2} + |\w|},
\ee
with 

\begin{align}
\G_J^* =\sum_{m}\G_{Jm}^*\;\; ;\;\; \G_{Jm}^*\equiv {1-2\eta_{m}^{-1} b_{Jm}^*\over\tau^{\star}_m},
\end{align}
and 
\begin{align}
&b_{Jm}^* = \\ &\int \frac{d\W_{\bs k} d\W_{\bs p}}{(4\pi)^2}\mrm{Tr}\left[Q_{m}(\hat{\bs{p}},\hat{\bs{k}},\alpha)F_{J}^\dag(\hat{\bs{k}})Q_{m}^{\mrm T}(-\hat{\bs{p}},-\hat{\bs{k}},\alpha)F_{J}(\hat{\bs{p}}) \right]. \nonumber
\end{align}
Interestingly, coefficients $b_{Jm}^*$ can be obtained from the corresponding zero-mass values $b_{Jm}$ from Table~\ref{Tbl:rhoc} (or The matrix~\ref{AppEq:bJm}) as

\begin{align}
b^*_{0gm} &= (1+I_m \alpha^2)b_{0gm}, \qquad s-\text{wave},  \\
b^*_{Jm} &= (1-\alpha^2) b_{Jm}, \qquad \text{non}-s-\text{wave} \, (J\ne 0g). \nonumber 
\end{align}
Consequently, we find for the pair-breaking rates $\Gamma^*_{Jm}$:

\begin{align} \label{Eq:Gamma*}
\Gamma^*_{0gm} &= \frac{1-2 b_{0gm}}{\tau_m^*} = \frac{(1-2b_{0gm})(1+I_m \alpha^2)}{\tau_m}, \\ \Gamma^*_{Jm} &= \frac{1-2(1-\alpha^2)\eta_m^{-1}b_{Jm}}{\tau_m^*} = \nonumber \\  &= \frac{1+I_m \alpha^2 - 2(1-\alpha^2) b_{Jm}}{\tau_m}, \qquad J\ne 0g, \nonumber   
\end{align}
where $\tau_m = \tau_m^* \eta_m$ is given by Eq.~\eqref{eq:tau} with $\nu_0$ replaced with the density of states $\nu_0^*$ for the massive Dirac spectrum. 

Equation~\eqref{Eq:Gamma*} immediately allows us to reproduce a number of well-known results. First, we see that Anderson's theorem holds~\cite{anderson1959theory}: time-reversal-invariant disorder ($m=0-5$) does not affect the $s$-wave channel ($J=0g$). Indeed, in this case all $b_{0gm} = 1/2$ leading to $\Gamma^*_{0gm} = 0$. We emphasize that this result holds for any mass $\delta$. Second, we can easily obtain the original result by Abrikosov and Gor'kov for $s$-wave superconductors with magnetic impurities~\cite{abrikosov1960contribution}. In this case, on has $b_{0gm}=-1/2$ for $m=6-15$, leading to $\Gamma^*_{0gm} \tau_m^* = 2$. Finally, in the limit of a single parabolic band without spin-orbit coupling, which formally corresponds to the case of an infinite mass $\alpha\to 1$, we recover the result by Larkin for $p$-wave pairing~\cite{Larkin1965}. In fact, if we consider any non-$s$-wave pairing ($J\ne 0g$) and any inversion-even type of disorder ($I_m=1$ for $m=0-1,\, 6-8, \, 13-15$), which includes density disorder, in the limit $\alpha \to 1$, we find that $\Gamma_{Jm}^* \tau^*_m = 1$, in agreement with Larkin. However, for inversion-odd disorder ($I_m=-1$ for $m=2-5, \, 9-12$), the result for any non-zero mass is not changed compared to the massless case, i.e., $\Gamma_{Jm}^* \tau^*_m = 1 - 2b_{Jm}$. Also, the  result $\Gamma_{2m}^* \tau^*_m = 1$ holds trivially for any mass in the channel $J=2$, since $b_{2m}=0$ for all $m$. Note, however, that $\tau_m^*$ itself strongly depends on mass and, as we discuss below, even diverges for inversion-odd disorder in the limit of an infinite mass, $\alpha \to 1$.

Another interesting observation that follows from Eq.~\eqref{Eq:Gamma*} is that any type of inversion-odd disorder ($I_m=-1$ for $m=2-5, \, 9-12$) does not affect superconductivity in the case of an infinite mass, $\alpha\to 1$. Indeed, all matrices $Q_m$ are independent of momentum in the limit $\alpha = 1$, implying that inversion-odd disorder only scatters between the conduction and valence bands. Such transitions are obviously suppressed  in the limit of large band gap, leading to the divergent single-particle scattering time $\tau_m^*\to\infty$ and, consequently, vanishing pair-breaking rate $\Gamma^*_{Jm}\to0$.

Finally, we comment on how finite mass affects the robustness of $J=0u$ channel. While finite mass breaks  $\mc{CT}$ symmetry and the channel becomes susceptible to most of the types of disorder, it is still unaffected by $\gamma_5$ ($m=2$) and $i \gamma_0 \gamma_5$ ($m=12$), which not only respect  $\mc{CT}$ symmetry, but also odd under inversion.

\section{Calculation of $T_{c}$}\label{app:Tc}

Using the condition for the superconducting instability, Eq.~\eqref{eq:instability} of the main text, we obtain that $\mc{P} \equiv -T_c\sum_{\omega}\mc{B}(i\omega) = -\dfrac{1}{g}$. 
Note that the derivation in this Appendix is independent of the pairing channel, so we omit the subscript $J$ for brevity.  
We now plug in the result for $\mc{B}(i\omega)$ as obtained in Eq.~\eqref{eq:BC2} and find
\begin{align}
\dfrac{1}{g} = T_c \sum_{n} \dfrac{\pi \nu_{0}}{\Gamma/2 +|\omega_n|}. \label{AppEq:1/g}
\end{align}
Using the definition of Matsubara frequencies $\omega_{n}=2\pi T (n+1/2)$, this equation can be rewritten as
\begin{align}\label{eq:SMc2}
\dfrac{2}{g\nu_{0}} = \sum_{n} \dfrac{1}{\Gamma/4\pi T_{c}+|n+1/2|}.
\end{align} 
Following Abrikosov and Gor'kov \cite{abrikosov1960contribution}, we make use of the fact that in the clean limit one has
\begin{align} \label{eq:clean}
\sum_{n\ge 0} \dfrac{1}{n+1/2}=\log \left(\dfrac{4 e^{\gamma_e}}{\pi} \dfrac{\omega_{D}}{2T_{c}} \right),
\end{align}
where $\gamma_e$ is the Euler's constant.
Thus, we can rewrite Eq.~\eqref{eq:SMc2} as
\begin{widetext}
\begin{align}
\begin{split}
\dfrac{1}{g\nu_{0}} = \sum_{n\ge 0} \dfrac{1}{\Gamma/4\pi T_{c}+(n+1/2)} = \sum_{n\ge0} \left[\dfrac{1}{\Gamma/4\pi T_{c}+(n+1/2)}-\dfrac{1}{n+1/2}\right]+\log \left(\dfrac{4 e^{\gamma_e}}{\pi} \dfrac{\omega_{D}}{2T_{c}} \right)\\
= \sum_{n\ge 0} \left[\dfrac{1}{(n+1/2+\Gamma/4\pi T_{c})}-\dfrac{1}{(n+1)}+\dfrac{1}{(n+1)}-\dfrac{1}{(n+1/2)}\right]+\log \left(\dfrac{4 e^{\gamma_e}}{\pi} \dfrac{\omega_{D}}{2T_{c}} \right).
\end{split}
\end{align}
\end{widetext}
We can identify the two terms inside the square brackets as digamma functions
\begin{align}
\Psi(z) = -\gamma_{e}+\sum_{n \ge 0} \left[\dfrac{1}{(n+1)}-\dfrac{1}{(n+z)}\right],
\end{align}
and we know from Eq.~\eqref{eq:clean} that $\dfrac{1}{g\nu_{0}}=\log \left(\dfrac{4 e^{\gamma_e}}{\pi} \dfrac{\omega_{D}}{2T_{c,0}} \right)$, where $T_{c,0}$ is the critical temperature for a clean system.
Consequently, we obtain
\begin{align}
\begin{split}
\log \left(\dfrac{4 e^{\gamma_e}}{\pi} \dfrac{\omega_{D}}{2T_{c,0}} \right) = \Psi(1/2)-\Psi(1/2+\Gamma/4\pi T_{c})+\\ + \log \left(\dfrac{4 e^{\gamma_e}}{\pi} \dfrac{\omega_{D}}{2T_{c}} \right),
\end{split}
\end{align}
or 
\begin{align}
\log \left(\dfrac{T_{c}}{T_{c,0}} \right) = \Psi(1/2)-\Psi(1/2+\Gamma/4\pi T_{c}),
\end{align}
which coincides with Eq.~\eqref{eq:criticalT} of the main text.

\section{Abrikosov-Gor'kov equations at arbitrary temperature: gapless superconductivity.}\label{app:gap}

Now we present the complementary approach to derive the effect of disorder on superconductivity, which exploits the formalism of Gor'kov Green's functions~\cite{Gorkov1958}. The advantage of this method is that it allows to treat the problem at arbitrary temperature and study thermodynamic and electromagnetic properties of a disordered superconductor at temperatures down to $T=0$. 

To start with, we introduce the Nambu space ($N$) for the MCBB electron operators according to
\be
\Psi_{\bs k} = \left( \begin{array}{c} c_{{\bs k}, 1} \\ c_{{\bs k}, 2} 
\\ c^\dagger_{ -{\bs k}, 1}\\ c^\dagger_{ -{\bs k}, 2} \end{array}\right)_N.
\ee

In this basis, the bare (without disorder) Gor'kov Green's function takes form~\cite{SigristUeda1991,MineevSamokhinbook}

\be 
\hat G_0(i\omega_n, {\bs k}) = - \frac{i \omega_n \hat \tau_0 + \xi_{\bs k} \hat \tau_3+ \hat \Delta_{\bs k}}{\omega_n^2 + \xi_{\bs k}^2 + \Delta_{\bs k}^2}, \label{AppEq:G0}
\ee 
where 
\be 
\hat \Delta_{\bs k} = \sqrt{2}\Delta\left( \begin{array}{cc} 0 & F^\dagger(\hat {\bs k}) \\ F(\hat {\bs k}) & 0  \end{array}\right)_{N}, \label{AppEq:hatDelta0}
\ee
and
\be
\Delta_{\bs k}^2 \equiv \Delta^2 {\Tr}\, F^\dagger(\hat {\bs k}) F(\hat {\bs k}).
\ee
Matrices $\hat \tau_0$ and $\hat \tau_3$ here are the corresponding Pauli matrices in the Nambu space (not to be confused with the Pauli matrices in the orbital basis). The factor $\sqrt{2}$ in Eq.~(\ref{AppEq:hatDelta0}) is introduced for convenience only, and simply reflects the normalization condition for functions $F(\hat {\bs k})$. The form of the Gor'kov Green's function~(\ref{AppEq:G0}) holds for the states with unitary pairing, i.e., satisfying the relation $\hat \Delta_{\bs k}^\dagger \hat \Delta_{\bs k}\propto \md 1$. The non-unitary states~\cite{SigristUeda1991}, which do not satisfy this relation and can be realized in multi-component superconductors, will be considered in future works.

The matrices $Q_m^{\a\b}({\bs p},{\bs k}) \equiv \langle \hat{\bs p}\a|M_m|\hat{\bs k}\b\rangle$, describing the scattering of electrons on the impurities of type $m$ in the MCBB basis, in the Nambu space become

\be 
Q_m(\hat{\bs p},\hat{\bs k}) \to \hat Q_m(\hat{\bs p},\hat{\bs k}) = \left( \begin{array}{cc} Q_m(\hat{\bs p},\hat{\bs k}) & 0 \\ 0 & - Q^T_m(-\hat{\bs k},-\hat{\bs p})    \end{array}   \right)_N.
\ee

The self-energy due to disorder is then given by 

\be  
\hat \Sigma_m (i\omega_n, \hat {\bs p}) = n_m V_m^2 \int \frac{d^3k}{(2\pi)^3}\hat Q_m(\hat{\bs p},\hat{\bs k}) \hat G(i \omega_n, \bs k) \hat Q_m(\hat{\bs k},\hat{\bs p}).
\ee
We notice that the self-consistency requires us to use full Green's function $\hat G$, instead of the bare one $\hat G_0$. Following Ref.~\cite{abrikosov1960contribution}, we look for a solution of the form

\be 
\hat G(i\omega_n, \bs {\bs k}) = - \frac{i \tilde \omega_n \hat \tau_0 + \xi_{\bs k} \hat \tau_3+ \tilde \Delta_{n, {\bs k}} }{\tilde \omega_n^2 + \xi_{\bs k}^2 + \tilde \Delta_{n, {\bs k}}^2},
\ee
with 
\be 
\tilde \Delta_{n, {\bs k}} = \sqrt{2}\tilde\Delta_n\left( \begin{array}{cc} 0 & F^\dagger(\hat {\bs k}) \\ F(\hat {\bs k}) & 0  \end{array}\right)_N,
\ee
and
\be
\tilde \Delta_{n, {\bs k}}^2 \equiv \tilde\Delta_n^2 {\Tr}\, F^\dagger(\hat {\bs k}) F(\hat {\bs k}).
\ee

Performing integration over $\xi_{\bs k}$ first, we obtain the very general expression which applies to any superconducting state with unitary pairing:

\begin{multline}
\hat \Sigma_m (i\omega_n, \hat {\bs p}) = - n_m V_m^2 \pi \nu_0 \int \frac{d\Omega_{\bs k}}{4 \pi}\hat Q_m(\hat{\bs p},\hat{\bs k}) \times \\ \times  \frac{i \tilde \omega_n \hat \tau_0 + \tilde \Delta_{n, {\bs k}}}{\sqrt{\tilde\omega^2_n + \tilde \Delta_{n, {\bs k}}^2 }}\hat Q_m(\hat{\bs k},\hat{\bs p}). \label{AppEq:self-energy}
\end{multline}

The above expression can be easily used to reproduce the result for the transition temperature $T_{c,J}$, Eq.~(\ref{eq:criticalT}). Neglecting $\tilde \Delta^2_{n,{\bs k}}$ in the denominator and performing integration over $\Omega_{\bs k}$ and summation over $m$, we find for the channel $J$

\begin{align}
&\hat \Sigma (i\omega_n, \hat {\bs p}) \equiv \sum_m \hat \Sigma_m (i\omega_n, \hat {\bs p})= \nonumber \\ & = -\frac{i \hat \tau_0 \sign  (\tilde \omega_{n})}{2\tau} + \frac{\tilde \Delta_{n,{\bs p}}(1-\tau \Gamma_J)}{2\tau|\tilde \omega_n|},
\end{align}
where $\Gamma_J$ is given by Eq.~\eqref{eq:Gamma}. In deriving the last equation, we also used Eqs.~(\ref{eq:amp}) and~(\ref{eq:b}). Utilizing further Dyson equation $\hat G^{-1} = \hat G_0^{-1} - \hat \Sigma$, we easily obtain

\begin{align}
&\tilde \omega_n = \omega_n + \frac{\sign (\tilde \omega_n)}{2\tau}, \nonumber \\
&\tilde \Delta_n = \Delta + \frac{\tilde \Delta_n (1- \tau \Gamma_J)}{2 \tau |\tilde \omega_n|},
\end{align}
which can be readily resolved yielding 
\begin{align}
&\tilde \omega_n = \omega_n + \frac{\sign (\omega_n)}{2\tau}, \nonumber \\
&\tilde \Delta_n = \Delta \left( 1 - \frac{1- \tau \Gamma_J}{2\tau |\tilde \omega_n|}  \right)^{-1}.
\end{align}

Finally, using the gap equation in the channel $J$
\be
\hat \Delta_{{\bs k}, \alpha \beta} = g_J T_c \sum_{n, {\bs p}} F_{J\alpha \beta}^\dagger (\hat {\bs k}) F_{J \gamma \delta} (\hat {\bs p}) \frac{\tilde \Delta_{n, {\bs p}, \delta \gamma}}{\tilde \omega_n^2 + \xi_{\bs p}^2},
\ee
we obtain after summation over $\bs p$

\begin{align}
1 &= \pi g_J T_c \nu_0 \sum_n \frac1{\Gamma_{J}/2+|\tilde \omega_n| - \frac1{2\tau}}= \nonumber \\ & = \pi g_J T_c \nu_0 \sum_n \frac1{\Gamma_{J}/2+|\omega_n|},
\end{align}
which is identical to Eq.~(\ref{AppEq:1/g}) and leads eventually to Eq.~(\ref{eq:criticalT}).

We emphasize that Eq.~(\ref{AppEq:self-energy}) is very general and can be used to study thermodynamic properties of any (unitary) superconducting state. As an example, we focus on the fully isotropic pairing functions $F_{0g}$ and $F_{0u}$. Performing integration over $\Omega_{\bs k}$ in Eq.~(\ref{AppEq:self-energy}), we find a set of coupled equations for $\tilde \omega_n$ and $\tilde \Delta_n$:

\begin{align} \label{Eq:tildeomegaDelta}
&\tilde \omega_n = \omega_n + \frac{\tilde \omega_n}{2\tau\sqrt{\tilde\omega_n^2 + \tilde \Delta_n^2}}, \nonumber\\ 
&\tilde \Delta_n = \Delta + \frac{\tilde \Delta_n(1-\tau \Gamma_J)}{2\tau\sqrt{\tilde\omega_n^2 + \tilde \Delta_n^2}},
\end{align}
accompanied with the gap equation

\be 
\Delta = \pi g_J T \nu_0 \sum_n \frac{\tilde \Delta_n}{\sqrt{\tilde\omega_n^2 + \tilde \Delta_n^2}}. \label{Eq:gapequationfull}
\ee
These equations, in principle, can be solved numerically to find the value of the pairing gap $\Delta$ at arbitrary temperature and study the thermodynamic properties of a superconductor. For instance, in the case of non-magnetic ($\mc{T}$-even) disorder for the $s$-wave $F_{0g}$ pairing or $\mc{CT}$-even disorder for the $p$-wave $F_{0u}$ pairing, we have $\Gamma_J=0$, and Eq.~\eqref{Eq:tildeomegaDelta} admits simple solution $\tilde \omega_n / \tilde \Delta_n = \omega_n / \Delta$. In these cases, the latter result implies that the gap equation~\eqref{Eq:gapequationfull} is not modified by disorder at all, consequently, the transition temperature and all the thermodynamic properties below $T_c$ remain unchanged compared to the clean case.

Abrikosov and Gor'kov have analyzed Eqs.~(\ref{Eq:tildeomegaDelta}) and~(\ref{Eq:gapequationfull}) in detail (with $\Gamma_J \ne 0$) for the most interesting limiting cases in Ref.~\cite{abrikosov1960contribution}. In particular, they found that there is a range of the impurity concentration where superconductivity is not entirely suppressed, while becoming gapless. In our language, this corresponds to the threshold value of the pair-breaking rate $\Gamma_{J}'$, above which the gap in the spectrum of elementary excitations vanishes:
\be 
\Gamma_{J}' = 2 e^{-\pi/4} \Gamma_{J \text{cr.}} \approx 0.91 \, \Gamma_{J \text{cr.}},
\ee 
where the critical value $\Gamma_{J \text{cr.}}$ is given by Eq.~(\ref{Eq:Gammacritical}). As a result, the low-temperature behavior of the specific heat changes from exponential to $T$-linear in the range $\Gamma'_{J} < \Gamma_J < \Gamma_{J\text{cr.}}$.

The analysis of this Appendix can be straightforwardly generalized to study the effect of different types of disorder on the anisotropic nematic pairing states in doped $\text{Bi}_2\text{Se}_3$ compounds~\cite{Funematic2014,VenderbosKoziiFu2016} or non-unitary chiral pairing in Majorana superconductors~\cite{VenderbosKoziiFu2016,KoziiVenderbosFu2016}. We leave these and related interesting questions to a future publication.

\section{Generalization of Anderson's argument to a generic anti-unitary symmetry}\label{app:AndersonCT}
In the main text we have seen that the $F_{0u}$ state is protected from any disorder respecting the product of time-reversal and chiral symmetries. In this Appendix we will show how to (trivially) generalize Anderson's original argument~\cite{anderson1959theory} to any antiunitary discrete symmetry that squares to minus one and acts within a {\it two-orbital} basis. 
To this end, we consider a discrete unitary symmetry $\mc C$, which acts within the space of the two orbitals, and construct the antiunitary symmetry by multiplying it by TRS, $\mc T$. We assume a basis of operators $\psi_{\tau\s}$, where $\tau = \pm$ is the orbital and $\s = \pm$  is spin (``$+$'' and ``$-$'' correspond to spin up and spin down, respectively), such that the symmetries act as follows
\begin{align}
&\mc T \psi_{\tau \s}(\bs k) = \s \psi_{\tau -\s}(-\bs k),\\
&\mc C \psi_{\tau \s}(\bs k) = \psi_{-\tau \s}(\bs k),
\end{align}
and thus,
\be
\mc{CT}\psi_{\s \tau}(\bs k) = \s \psi_{-\tau -\s}(-\bs k).
\ee
For the sake of brevity we will denote orbit and spin under the same index $\a = (\s,\tau)$, such that 
\be
\mc{CT}\psi_{\bs k \a} = \a \psi_{-\bs k-\a},
\ee
where the convention is that $-\a$ has both spin and orbit flipped compared to $\a$ and that the sign of $\a$ is given only by the spin component, such that $\a = 1$ for $(+,\tau)$ and $\a = -1$ for $(-,\tau)$.

We will now show that the transition temperature of the pairing state 
\be \label{eq:app:pairing_state}
\D ={1\over 2} \psi_{\bs k} \,\mc{CT}\, \psi_{\bs k}  = \sum_{\a}{\a\over 2}\psi_{\bs k\a}\psi_{-\bs k-\a}
\ee
is not affected by disorder that respects the product $\mc{CT}$. 
Notice that when $\mc C$ anticommutes with inversion this state is an odd-parity pairing state. 
The pairing state~\eqref{eq:app:pairing_state} is driven by the interaction Hamiltonian  
\begin{align}\label{eq:app:H_I}
\mc H_I &= -g \sum_{\bs k \bs p}\psi_{\bs p}^{\dag} \,(\mc{CT})^{-1} \, \psi_{\bs p}^\dag \;\psi_{\bs k} \,\mc{ C T }\, \psi_{\bs k} \\
&= -g  \sum_{\alpha \beta} \sum_{\bs k \bs p}\a\b\,\psi_{-\bs p-\a}^{\dag}  \psi_{\bs p\a}^\dag \;\psi_{\bs k\b}  \psi_{-\bs k-\b}.\nn
\end{align}

To show that the pairing state is not affected by disorder we follow Anderson's original argument~\cite{anderson1959theory}. We consider a generic dispersion Hamiltonian 
\be 
\mc H_0 = \sum_{\bs k}\psi^\dag _{\bs k} \hat h_{\bs k} \psi_{\bs k} \label{AppEq:H0}
\ee
 and a disorder Hamiltonian 
\be 
\mc H_d  = \sum_{\bs k \bs p} \psi_{\bs p}^\dag \hat d_{\bs p \bs k} \psi_{\bs k}, \label{AppEq:Hd}
\ee
where $\hat d_{\bs p \bs k}=\sum_{lm} e^{i(\bs k-\bs p)\cdot \bs r_l} M_m $.
It is assumed that both Hamiltonians~\eqref{AppEq:H0} and~\eqref{AppEq:Hd} respect $\mc{CT}$. Notice that the disorder potential or dispersion Hamiltonian need not respect $\mc T$ or $\mc C$ individually, but only the product of the two. 

Following the Anderson argument, we now diagonalize the sum of dispersion and disorder Hamiltonians  
\be
\sum_{\bs p \bs k \a\b}[U_{\bs n \bs p}^{a \a}]^* (\hat h_{\bs p}^{\alpha \beta} \delta_{\bs {k p}} +\hat d_{\bs p\bs k}^{\alpha \beta}) U_{\bs n' \bs k}^{b \b} = \e_{\bs n,a} \delta_{\bs n \bs n'}\delta_{a b}
\ee
and the corresponding field operators 
\be\label{eq:app:c}
c_{\bs n a}=[U_{\bs n \bs p}^{a\a}]^* \psi_{\bs p \a}\,,
\ee
where $\bs n$ denotes the spatial states that diagonalize the above Hamiltonian. Because the sum $\mc H_0 + \mc H_d$ possesses $\mc{CT}$ symmetry we may assume that the new operators $c_{\bs n a}$ come in pairs related to each other under $\mc{CT}$. Thus we can denote the $\mc{CT}$ partner of $c_{\bs n a}$ by $c_{-\bs n -a}$ (given the symmetry, we can always label states in this manner). It then follows that 
\begin{align}\label{eq:app:sym}
&(\mc{CT})^{-1} U_{\bs n \bs k}^{a\a} \mc{CT}= a\a[U_{-\bs n,-\bs k }^{-a-\a}]^* = U_{\bs n \bs k}^{a\a}, 
\end{align}
where, as before, the sign of the orbital indices $a$ and $\a$ is given by the spin component alone. The first equality in Eq.~\eqref{eq:app:sym} stems from the definition of $\mc {CT}$ symmetry, while the second one is true because the Hamiltonian preserves this symmetry.  

\begin{widetext}
We are now ready for the final step. We transform the operators in Eq.~\eqref{eq:app:H_I} to the basis that diagonalizes the sum of $\mc H_0+\mc H_d$ according to Eq.~\eqref{eq:app:c}:
\begin{align}
\mc H_I &= -g \sum_{\bs k \bs p}\sum_{\bs n_1\bs n_2\bs n_3\bs n_4}\sum_{\a \b a b e l} \a\b \left\{ c_{\bs n_1 e}^{\dag}[U_{\bs n_1 -\bs p}^{e -\a}]^*  [U_{\bs n_2 \bs p }^{l \a  }]^*  c_{\bs n_2 l}^\dag\right\} \left\{ c_{\bs n_3 b} U_{
\bs n_3 \bs k }^{b\b}  U_{\bs n_4    -\bs k }^{a    -\b }c_{\bs n_4 a} \right\}\, \\
&= -g \sum_{\bs k \bs p}\sum_{\bs n_1\bs n_2\bs n_3\bs n_4}\sum_{\a \b a b e l} a e \left\{ c_{\bs n_1 e}^{\dag}U_{-\bs n_1 \bs p}^{-e \a}  [U_{ \bs n_2 \bs p}^{l \a }]^*  c_{\bs n_2 l}^\dag\right\} \left\{ c_{\bs n_3 b} U_{
\bs n_3 \bs k }^{b\b}  [U_{-\bs n_4 \bs k }^{-a \b }]^*c_{\bs n_4 a} \right\}\,\nn
 \\
&= -g \sum_{\bs n \bs n'}\sum_{a e} a e \,c_{-\bs n -e}^\dag c_{\bs n e}^\dag \,c_{\bs n' a}c_{-\bs n' -a}\nn,
\end{align}
where in the second line we used Eq.~\eqref{eq:app:sym}. The final result is that this Hamiltonian has exactly same form as Eq.~\eqref{eq:app:H_I} and the interaction weight $g$ remains  unchanged in the new basis. Consequently, $T_c$ is not modified as long as the density of the energy states $\e_{\bs n a}$ is the same as in the clean Hamiltonian.  

\section{The conversion matrix $b_{Jm}$}\label{app:reps}

The matrix elements $b_{Jm}$ from Eq.~\eqref{eq:b}, where  $m=0-15$ numerates different types of disorder, equal to

\be
b_{Jm} = 
\begin{pmatrix} \label{AppEq:bJm}
{1/ 2} & {1/ 2}   & {1/ 2}  & {1/ 2} & {1/ 2} & {1/ 2} & -{1/ 2}  & -{1/ 2}  & -{1/ 2}  &-{1/ 2}  & -{1/ 2} & -{1/ 2} & -{1/ 2} &-{1/ 2} & -{1/ 2} & -{1/ 2}
\\

{1/ 2} & -{1/ 2} & {1/ 2} &  -{1/ 2} &-{1/ 2} &-{1/ 2} & -{1/ 2}  &-{1/ 2}  & -{1/ 2} &-{1/ 2}  & -{1/ 2} &-{1/ 2} &{1/ 2}   &{1/ 2}   &{1/ 2}   & {1/ 2}
\\

{1/ 3} & -{1/ 3}  &-{1/ 3}  &{1/ 3}  &-{1/ 3} &-{1/ 3} &-{1/ 3}  & {1/ 3}   &{1/ 3}   &{1/ 3}   &-{1/ 3}  &-{1/ 3} &-{1/ 3} &{1/ 3}   &-{1/ 3}   &-{1/ 3}  
\\

{1/ 3} &-{1/ 3}   &-{1/ 3}  &-{1/ 3} &{1/ 3}  &-{1/ 3} & {1/ 3}   & -{1/ 3}  & {1/ 3}   &-{1/ 3}  &{1/ 3}  &-{1/ 3} &-{1/ 3}  &-{1/ 3}  &{1/ 3}   & -{1/ 3}  
\\

{1/ 3} & -{1/ 3}  &-{1/ 3}  &-{1/ 3} &-{1/ 3} &{1/ 3}  & {1/ 3}   & {1/ 3}   & -{1/ 3}  &-{1/ 3}  & -{1/ 3} &{1/ 3}  & -{1/ 3} &-{1/ 3}  &-{1/ 3}  &{1/ 3} 
\\

0   &0      &0     &0    &0    &0    & 0     & 0     & 0     &0     &0     &0    &0     &0     &0     & 0
\end{pmatrix}.
\ee
The last line of this matrix describes {\it all} channels with $J=2$ [i.e., $F_{21}(\hat {\bs k})-F_{25}(\hat {\bs k})$].

\end{widetext}

\end{document}